\documentclass[aps,amsfonts,amsmath,prd,preprint,nofootinbib]{revtex4}
\pdfoutput=1
\newcommand{\beq}{\begin{equation}}
\newcommand{\eeq}{\end{equation}}
\usepackage{graphicx}
\usepackage{natbib}
\usepackage{xcolor}

\usepackage{amsmath}

\usepackage[utf8]{inputenc} 
\usepackage{graphicx,xcolor,overpic,mathtools}
\usepackage[colorlinks]{hyperref}
\hypersetup{ 
	colorlinks=true, 
	linkcolor=black, 
	filecolor=black, 
	citecolor = blue,       
	urlcolor=blue, 
} 

\newcommand{\bea}{\begin{eqnarray}}
\newcommand{\eea}{\end{eqnarray}}
\newcommand{\bi}{\begin{itemize}}
\newcommand{\ei}{\end{itemize}}

\usepackage{upgreek}
\usepackage{csquotes}
\numberwithin{equation}{section}

\usepackage[title]{appendix}
\usepackage{verbatim}
\usepackage{tensor}
\usepackage{tikz}
\usepackage[compat=1.1.0]{tikz-feynman}

\def\beq{\begin{equation}} 
\def\eeq{\end{equation}}
\def\beqa{\begin{eqnarray}}
\def\eeqa{\end{eqnarray}}

\usepackage{amsthm,amsmath,amssymb}

\PassOptionsToPackage{normalem}{ulem}
\usepackage{ulem} 
\usepackage{sidenotes}
\usepackage{bm}
\usepackage{url}
\usepackage{physics}
\usepackage{xfrac}
\usepackage[makeroom]{cancel}
\usepackage{tensor}
\usepackage{mathrsfs}
\usepackage{slashed}
\usepackage{tikz}
\usepackage[mode=buildnew]{standalone}
\usepackage{bbm}
\usepackage{booktabs}
\usepackage{cleveref}
\usepackage{placeins}
\usepackage{ragged2e}
\crefname{section}{Sec.}{sec.}
\crefname{figure}{Fig.}{Fig.}
\Crefname{section}{Section}{Sections}
\usepackage[caption=false]{subfig}

\NewDocumentCommand{\evat}{sO{\bigg}mm}{%
  \IfBooleanTF{#1}
   {\mleft. #3 \mright|_{#4}}
   {#3#2|_{#4}}%
}

\newtheorem{theorem}{Theorem}

\begin{document}

\title{Localized Vector Modes on Local Cosmic Strings}

\author{Sergio Alameda-Calvo{$^{1}$}\footnote{sergio.alameda@usal.es}, Jose J. Blanco-Pillado{$^{2,3,4}$}\footnote{josejuan.blanco@ehu.eus} and Jose Queiruga{$^{1,5}$}\footnote{xose.queiruga@usal.es}}

\affiliation{
$^1$
Institute of Fundamental Physics and Mathematics,
University of Salamanca, 37008 Salamanca, Spain\\
$^2$ Department of Physics, University of Basque Country, UPV/EHU, 48080, Bilbao, Spain \\
$^3$ EHU Quantum Center, University of Basque Country, UPV/EHU, 48080, Bilbao, Spain \\
$^4$ IKERBASQUE, Basque Foundation for Science, 48011, Bilbao, Spain \\
$^5$ Department of Applied Mathematics,
University of Salamanca, 37008, Salamanca, Spain.
}

\begin{abstract}

We investigate the dynamics of vector excitations localized on cosmic strings in the Abelian-Higgs model. These confined gauge field fluctuations behave as massive vector degrees of freedom propagating along the string. We show that they remain bounded for all values of the scalar self coupling $\lambda$, including parameter regimes in which the previously studied conventional {\it shape mode} is no longer present. After determining their spectra and spatial profiles, we analyze their decay through nonlinear coupling to bulk radiation. Their amplitudes exhibit the characteristic power law relaxation associated with non-linear radiation. For sufficiently large $\lambda$ the scalar radiation channel becomes kinematically inaccessible and the decay is instead mediated by quasinormal modes which can be interpreted as Feshbach resonances of the string. We then study the full $3+1$ dimensional dynamics of the vector-excited string. Field theory simulations reveal a parametric instability between the vector mode and the Goldstone sector. In contrast to the shape mode, the vector mode resonantly excites both transverse directions simultaneously. In addition, we build an effective model that captures the instability and identifies the nonlinear couplings responsible for it. Our results show that vector excitations constitute a long-lived massive degree of freedom with a non-trivial role in the dynamics of local strings.

\end{abstract}

\maketitle

\section{Introduction}

Cosmic strings are one dimensional topological defects that may have formed during symmetry breaking phase transitions in the Early Universe \cite{Kibble:1976sj}. Their formation, evolution, and potential observational consequences have been studied extensively over the past several decades \cite{Vilenkin:2000jqa}.

Many of the observable signatures of cosmic string networks depend sensitively on their dynamical evolution. A reliable assessment of these signatures therefore requires a detailed understanding of string dynamics over extremely long, potentially cosmological, timescales. Among the most promising observational probes are the gravitational wave signals generated by oscillating strings and string loops. Such signals constitute an important cosmological target for both current and future gravitational wave observatories \cite{LIGOScientific:2021nrg,Blanco-Pillado:2024aca,NANOGrav:2023hvm,Auclair:2019wcv,ET:2025xjr}. A precise characterization of cosmic string dynamics under different physical conditions is therefore essential for robustly connecting gravitational wave observations to the underlying high energy theories responsible for string formation. In particular, uncertainties in the microscopic physics of strings can affect the predicted gravitational wave spectrum and thereby influence the constraints inferred on the fundamental parameters of the underlying model.

On the other hand, direct field theory simulations of cosmic strings are, however, severely constrained by the enormous hierarchy between the microscopic width of the string core and the macroscopic length scales relevant to their cosmological evolution. This separation of scales naturally motivates the development of effective theories describing the low-energy, long-wavelength dynamics of strings. At leading order, this dynamics is governed by the Nambu-Goto action, which describes the motion of an infinitely thin relativistic string \cite{Goto:1971ce,nambu1970lectures}. Its propagating degrees of freedom are the translational Goldstone modes associated with the spontaneous breaking of spatial translations transverse to the string. In the linearized spectrum, these appear as zero modes corresponding to rigid displacements of the string core.

The Nambu–Goto description may, in principle, be systematically extended by incorporating higher order operators that encode finite width effects. In particular, curvature dependent terms capture deviations from the infinitely thin string approximation. Such corrections have been studied extensively in the past \cite{Maeda:1987pd,Gregory:1988qv,Anderson:1997ip} and were recently revisited in Ref.~\cite{Aurrekoetxea:2026xrl}, where it was shown that the leading contributions of this type vanish for Abelian–Higgs strings.

A qualitatively distinct extension arises from the possible existence of massive degrees of freedom localized on and propagating along the string. The role of such internal modes in soliton dynamics has attracted considerable attention in recent years. The excitation and radiative decay of solitons in (1+1) dimensions are by now well understood in a variety of situations; see, for example, Ref.~\cite{Manton:2004tk,Manton:1996ex, Blanco-Pillado:2020smt}. Internal modes are also known to play a central role in soliton scattering, giving rise to resonant energy transfer and intricate interaction phenomena \cite{Sugiyama:1979mi,Campbell:1983xu}. By contrast, considerably less is known about their influence on extended, stringlike defects. This question has therefore become an active area of research, encompassing domain wall strings \cite{Blanco-Pillado:2022rad,Blanco-Pillado:2024bev}, global strings \cite{Blanco-Pillado:2022axf}, and local strings \cite{Aurrekoetxea:2026xrl}.

Internal excitations can likewise have important consequences for vortex interactions, where the presence of localized modes within the vortex cores may generate qualitatively new collective phenomena \cite{Alonso-Izquierdo:2024fpw,Bachmaier:2025igf,Alonso-Izquierdo:2025suz,AlonsoIzquierdo:2026mub}. More broadly, the excitation and subsequent decay of massive string modes have been proposed as a possible mechanism for reconciling the apparent differences between field theory simulations of cosmic string networks \cite{Hindmarsh:2021mnl} and their Nambu–Goto counterparts \cite{Blanco-Pillado:2013qja}.

In the context of local strings, Ref.~\cite{Aurrekoetxea:2026xrl} presented a detailed analysis of the so called shape mode (a scalar mode made of a combination of the underlying scalar and vector $3+1$ dimensional fields) and its coupling to the translational Goldstone sector. The effective interactions responsible for this coupling were identified, and their influence on the dynamics of both straight strings and closed loops was investigated in the context of field theory simulations.

In addition to this massive scalar excitation, local strings may support localized vector modes. These excitations arise in the linearized fluctuation spectrum and are characterized by nonvanishing perturbations confined entirely to the gauge field sector.

In this work, we extend and complete earlier investigations of localized vector modes on Abelian–Higgs strings \cite{Davis:1987df,Arodz:1991ws,Arodz:1995pt,Arodz:1996eb} by determining their spatial profiles, energies, nonlinear interactions, and decay mechanisms. In contrast to the shape mode, which disappears into the continuum above a critical value of the scalar self-coupling ($\lambda$), the vector mode remains localized and does not appear to merge with the continuum. Its lifetime is comparable to that of the shape mode. Generically, the vector excitation couples quadratically to radiative degrees of freedom, leading to an asymptotic amplitude decay proportional to ($t^{-1/2}$). We derive this decay law using a perturbative framework analogous to that introduced in Ref.~\cite{Manton:1996ex}. For sufficiently large values of ($\lambda$), however, we show that the vector mode can excite quasinormal modes of the string \cite{Alonso-Izquierdo:2024tjc}, thereby opening additional decay channels. 

We further demonstrate that vector mode excitations can trigger a parametric instability that transfers energy to the translational zero mode sector. We characterize the nature of this instability and highlight its qualitative differences from the instability generated by shape mode excitations. The coexistence of these mechanisms suggests a rich interplay among internal massive modes, translational degrees of freedom, and radiative channels in generically excited strings.

The remainder of this paper is organized as follows. In Sec.~\ref{model}, we introduce the Abelian–Higgs model, review the static vortex solution, and examine the internal excitations supported by the string, with particular emphasis on the characterization of the vector modes. In Sec.~\ref{sec:rad}, we investigate the decay mechanisms of the vector excitation. Section~\ref{Sec: interac} is devoted to its coupling to the transverse degrees of freedom of the string and to the resulting parametric instability. Finally, in Sec.~\ref{conclusions}, we summarize our main results and discuss their broader implications. Technical details of several calculations are provided in Appendices~\ref{App.1}--\ref{App:5}.


\section{The Model.  Conventions and spectrum.}
\label{model}

We will consider the Abelian-Higgs model in 3+1 dimensions. In dimensionless variables, the model is defined by the following Lagrangian density
\begin{equation}
    \mathcal{L}=-\frac{1}{4}F_{\mu\nu}F^{\mu\nu}+\frac{1}{2}(D_\mu\phi)^*D^\mu\phi-\frac{\lambda}{8}(1-\phi^*\phi)^2\,. \label{dimless_lag}
\end{equation}

The covariant derivative for the complex scalar field $\phi$ is defined as $ D_\mu\equiv\partial_\mu-iA_\mu$ and $F_{\mu\nu}=\partial_\mu A_\nu-\partial_\nu A_\mu$ is the field strength tensor for the gauge field $A_\mu$. The field equations are
\begin{align}\label{EOM_phi}
    D_\mu D^\mu\phi-\frac{\lambda}{2}(1-\phi^*\phi)\phi&=0\,,\\
    \partial_\mu F^{\mu\nu}+\frac{i}{2}\left[\phi^*D^\nu\phi-\phi(D^\nu\phi)^*\right]&=0\,.\label{EOM_A}
\end{align}

From now on we will work in the temporal gauge, $A_0=0$. The Abrikosov-Nielsen-Olesen (unit charge) vortex string is a static and topologically non-trivial solution to the previous field equations \cite{Abrikosov:1956sx, Nielsen:1973cs}. In cylindrical coordinates, it is given by

\begin{equation}
    \phi^v(r,\theta)=f(r)e^{i\theta}\,,\quad A^v_r=0\,,\quad A^v_\theta=\frac{a(r)}{r}\,,\quad A_z^v=0\,,
\end{equation}
where the profile functions $f(r)$ and $a(r)$ fulfill
\begin{align}
    f''+\frac{f'}{r}-\frac{(1-a)^2}{r^2}f+\frac{\lambda}{2}(1-f^2)f=\,&0\,,\\
    a''-\frac{a'}{r}+(1-a)f^2=\,&0\,.
\end{align}

\begin{figure}[h]
    \centering
    \includegraphics[width=\textwidth]{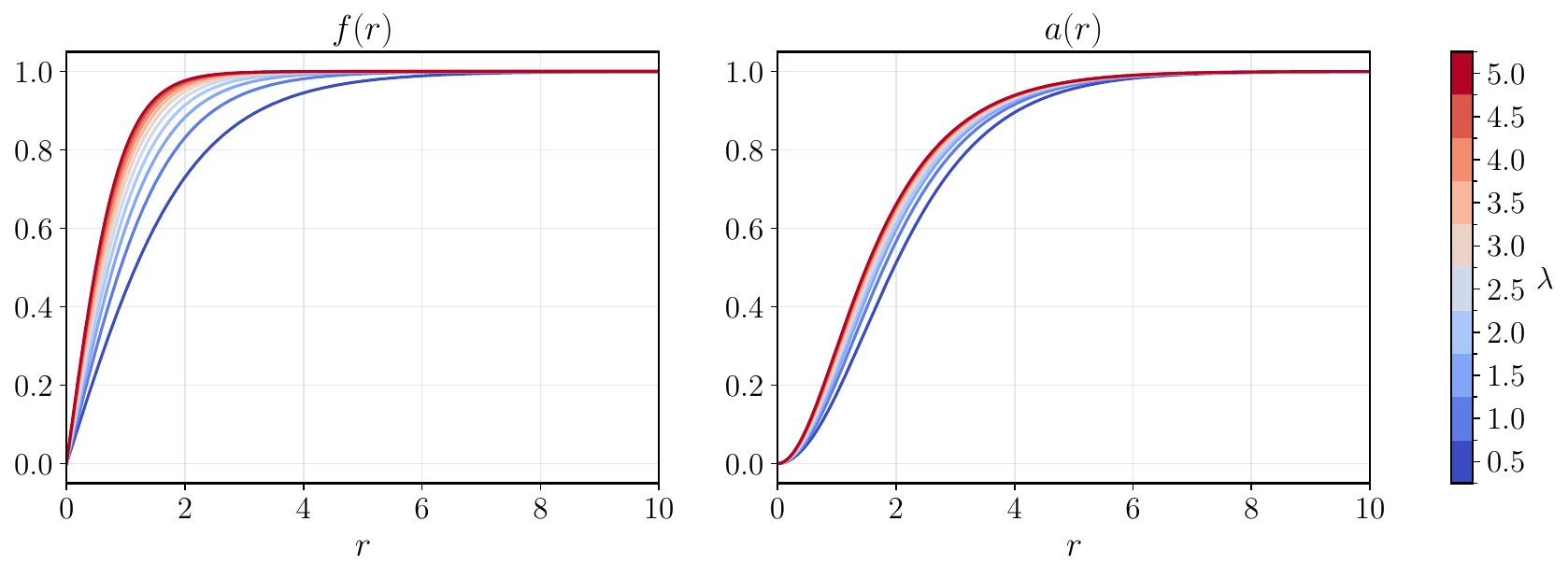}
    \caption{Profile functions $f(r)$ and $a(r)$ of the vortex solution for several values of the coupling constant, $\lambda$.}
    \label{fig:vortex_profiles}
\end{figure}

Here, primes denote differentiation with respect to the radial coordinate, $r$. Moreover, the profile functions are complemented by the boundary conditions, $f(0)=a(0)=0$ and $f(\infty)=a(\infty)=1$.  In Fig. \ref{fig:vortex_profiles} we show the static vortex profiles for different coupling constants.

The vortex string possesses two characteristic core widths, set by the inverse of the masses of the corresponding vacuum excitations, $\delta_\phi\sim m_\phi^{-1}$ and $\delta_A\sim m_A^{-1}$, with $m_\phi=\sqrt{\lambda}$ and $m_A=1$, in dimensionless units.

In the local theory, the vortex string carries an intrinsic magnetic field along its axis
\begin{equation}
    B_z^v(r)=\frac{1}{r}F_{r\theta}=\frac{a'(r)}{r},
\end{equation}
which gives rise to a classically quantized total magnetic flux
\begin{equation}
    \Phi_B=\int_{\mathbb{R}^2}\mathbf{B}\cdot d\mathbf{S}=2\pi n\,,
\end{equation}
$n$ being the winding number, or topological charge, of the vortex (In the following we will take $n=1$.). In this regard, the local string can be seen as a straight magnetic flux tube.

Local vortices in the Abelian-Higgs model, and in particular, the role of their internal excitations, have been studied in the literature in a variety of dynamical settings, including vortex-antivortex collisions \cite{Bachmaier:2025igf}, spectral-walls \cite{Alonso-Izquierdo:2024fpw}, and modifications of moduli-space dynamics induced by the presence of these excitations \cite{AlonsoIzquierdo:2026mub, Alonso-Izquierdo:2025suz}, among others. Although the present work is mainly devoted to the study of gauge perturbations of the string, other types of internal modes will have some relevance in what follows. We therefore first provide a brief review of some of the relevant modes already discussed in the literature.

Let us begin with translational zero modes. The translational mode is the only mode with zero eigenvalue. Whereas in the global theory they are generated by ordinary derivatives of the background fields along transverse directions, in the local theory that translation must be supplemented by an appropriate gauge transformation. Thus, in the $j$-direction it can be obtained from the following gauge fixed translations
 \begin{align}
    \phi(x)=\phi^v(x)+D_j\phi(x)\cos(k z)\, \delta x^j,\\
    A_i(x)=A^v_i(x)+F_{ij}(x)\cos(k z)\,\delta x^j.
\end{align}
Note that for $k=0$ they correspond to rigid translations of the string, while for $k\neq 0$ they can be interpreted as wave-like deformations in the $z$-direction.

We next turn to the scalar shape mode, which has been studied before in the literature; see, for example, Ref.~\cite{Goodband:1995rt,Kojo:2007bk,Alonso-Izquierdo:2015tta,Alonso-Izquierdo:2016pcq}. In the following, we closely follow the analysis presented in Ref.~\cite{Alonso-Izquierdo:2024tjc}.
In the temporal gauge, the shape mode can be obtained as a perturbation of the scalar field and of the angular component of the gauge field, namely,
\begin{align}\label{ansatz_shape1}
    \phi(x)=f(r)e^{i\theta}+\epsilon\, \varphi(r)e^{i\theta}e^{i\left(\omega t-k z\right)},\\ \label{ansatz_shape2}
    A_\theta(x)=\frac{a(r)}{r}+\epsilon \alpha(r)e^{i\left(\omega t-kz\right)}.
\end{align}
 
 This perturbation leads to a two-dimensional Sturm-Liouville problem of the form
 \begin{equation}\label{eq: shape_mode_spec_problem}
     \mathcal{H}\begin{pmatrix}
\varphi_n(r) \\
\alpha_n(r)
\end{pmatrix}=\Omega_n^2\begin{pmatrix}
\varphi_n(r) \\
\alpha_n(r)
\end{pmatrix} \,, 
\quad \left(\mathcal{H}\right)_{ij}=-\delta_{ij}\left(\frac{d^2}{dr^2}+\frac{1}{r}\frac{d}{dr}\right)+V_{ij}(r)\,,
 \end{equation}
with
 \begin{align}
     &V_{\phi\phi}(r)=\frac{3}{2}\lambda f(r)^2-\frac{\lambda}{2}+\frac{(1-a(r))^2}{r^2}\,,\\
     &V_{\phi\theta}(r)=V_{\theta\phi}(r)=-\frac{2f(r)}{r}(1-a(r))\,,\\
      &V_{\theta\theta}(r)=f(r)^2+\frac{1}{r^2}\,,\\
      & \Omega_n^2=\omega_n^2-k_n^2\,,
 \end{align}
whose square integrable solutions cease to exist for $\Omega_n>1$ at $\lambda\approx 1.5$, see \cite{Alonso-Izquierdo:2024tjc} for details. In Fig. \ref{fig:spectrum} we show the frequency of the homogeneous mode as a function of $\lambda$.  

For $\lambda\gtrsim 1.5$, there are no genuine bound modes that respect the symmetry induced by perturbation (\ref{ansatz_shape1})-(\ref{ansatz_shape2}). The underlying reason is straightforward: at $\lambda \sim 1.5$, the mode frequency reaches $\Omega = 1$, which coincides in our units with the mass threshold of the vector component of the mode $m_A$. As a consequence, the vector component becomes a scattering state.

However, the mass threshold of the scalar component, $m_\phi$, lies above this value, and therefore the scalar component can not propagate. In this way, within the spectral interval $(1, \sqrt{\lambda})$, there exist modes whose scalar component behaves as a bound mode, while their vector component is a scattering mode. These modes are referred to as quasinormal (or quasi-bound) modes \cite{Alonso-Izquierdo:2024tjc}. Although they decay at linear order, they are similar to genuine bound modes in the sense that they can have long lifetimes and interact with other modes. We will see later that they play a fundamental role in the decay of vector modes in the large $\lambda$ regime, in the type II strings.

\subsection{Vector modes}
 
\begin{figure}[h]
    \centering
    \includegraphics[width=\textwidth]{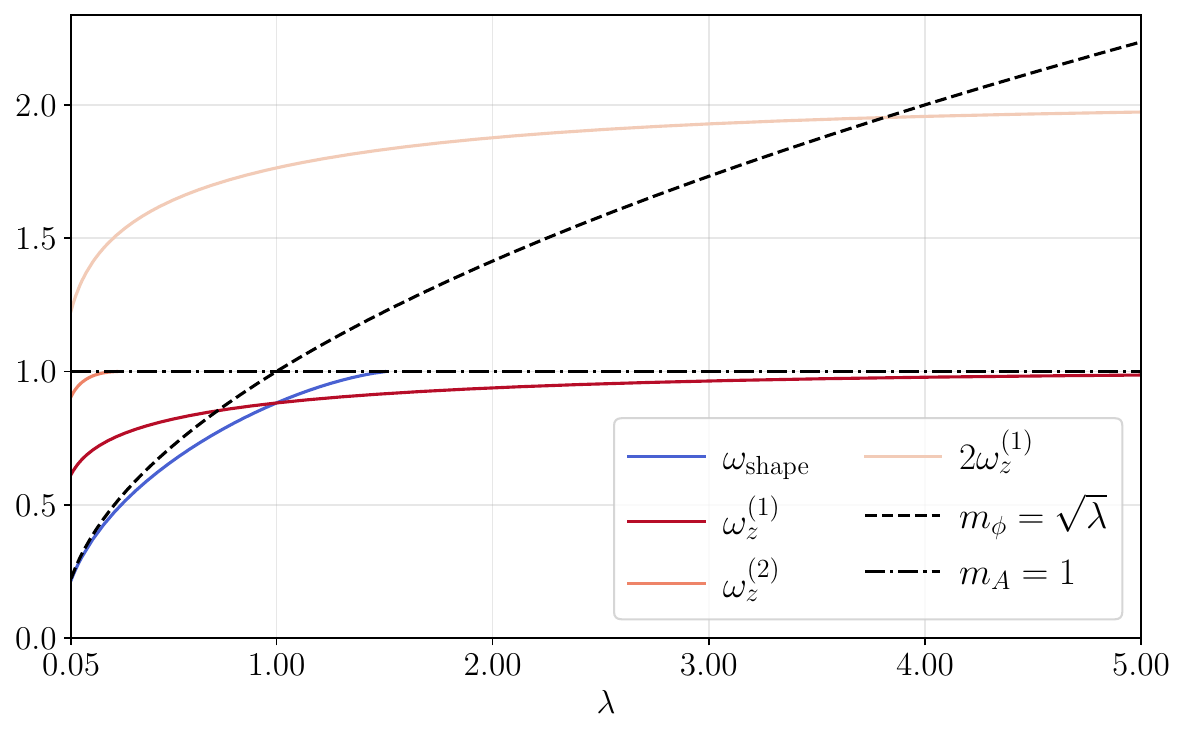}
    \captionsetup{justification=centering}
    \caption{Spectrum of the vector and shape modes as a function of $\lambda$.}
    \label{fig:spectrum}
\end{figure}

In addition to these modes, the Abelian–Higgs string also supports an axially symmetric, massive bound state associated to the $z$-component of the gauge field \cite{Davis:1987df,Arodz:1991ws, Arodz:1995pt, Arodz:1996eb}. Contrary to the previous ones, this mode is intrinsic to the three-dimensional string. Throughout the work, we will refer to this bound state as the \textit{vector mode}. Let us consider the following perturbation
 \begin{align}\label{eq:vector_pert1}
    \phi(x)&=\phi^v(x),\\ \label{eq:vector_pert2}
    A_i(x)&=A^v_i(x),\,\quad i=x,y\\ \label{eq:vector_pert3}
    A_z(x,t)&=A_z^v(x)+\epsilon\Psi(x,t)~.
\end{align}

The vector mode is a solution to the Schrödinger-like equation that arises from the field equations (\ref{EOM_phi}) and (\ref{EOM_A}) in the string background with the perturbations (\ref{eq:vector_pert1})-(\ref{eq:vector_pert3}), namely
\begin{align}\label{eq:vec_sh1}
-\ddot f+f''+\frac{f'}{r}-\left(\frac{(1-a^2)}{r^2}+\Psi^2\right)f+\frac{\lambda}{2}(1-f^2)f&=0\,,\\ \label{eq:vec_sh2}
-\ddot{a}+a''-\frac{a'}{r}+(1-a)f^2&=0\,,\\ \label{eq:vec_sh3}
    -\ddot \Psi+\Psi''+\frac{\Psi'}{r}-\Psi f^2&=0\,.
\end{align}

Note that equations (\ref{eq:vec_sh1}) and (\ref{eq:vec_sh2}) are satisfied at linear order in $\epsilon$ for the static vortex solution. That is, the vector mode is a solution to a single Schrödinger-like equation given by (\ref{eq:vec_sh3}). 
Assuming an oscillatory ansatz for the fluctuations of the form 
\begin{equation}
\Psi(r,t)=\psi(r)e^{i\omega_z t}~,
\end{equation}
the eigenvalue equation for the radial profile reads
\begin{equation}\label{eq: eigenproblem vector mode}
    -\psi''(r)-\frac{\psi'(r)}{r}+\psi(r) f(r)^2=\omega_z^2\psi(r)\,.
\end{equation}

\begin{figure}[h]
    \centering
    \includegraphics[width=0.8\textwidth]{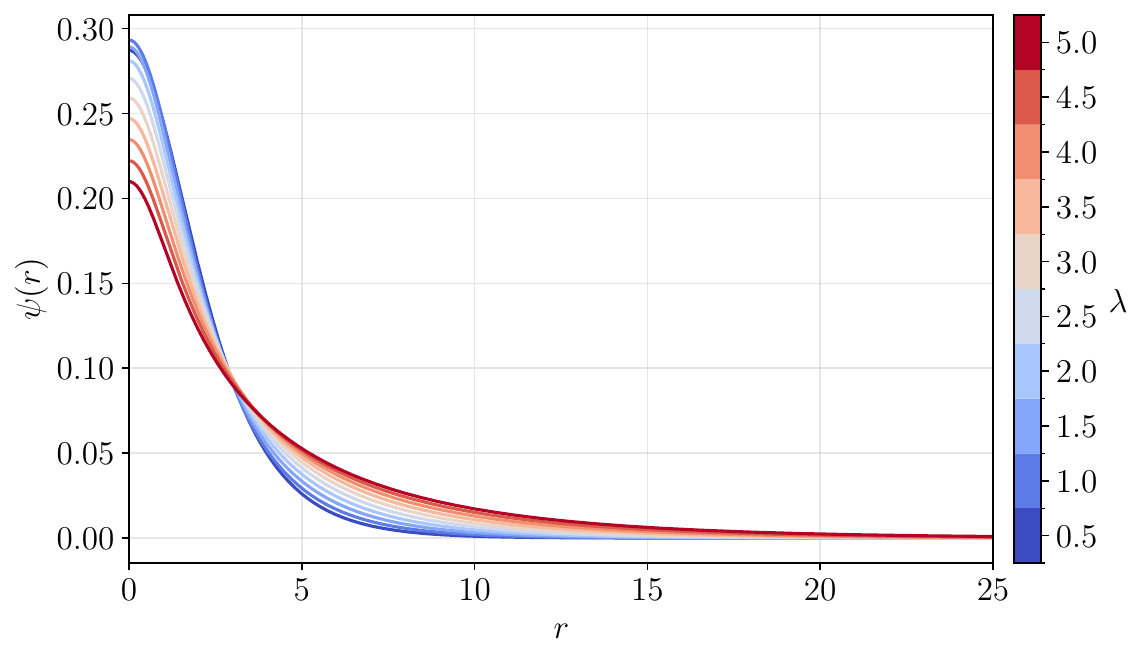}
    \caption{\justifying
         Radial profiles of the normalized vector mode for different $\lambda$ values. The modes are normalized to the standard $L^2$ norm on the plane.}
    \label{fig: vector mode profiles}
\end{figure}

By numerically solving this eigenvalue problem, one gets the spectrum for the discrete mode in terms of  $\lambda$ as well as the radial profile $\psi(r)$ shown in \cref{fig:spectrum} and \cref{fig: vector mode profiles}, respectively.

Unlike the shape mode, the vector mode never merges into the continuum, remaining a bound state for arbitrarily large $\lambda$. This follows from a classical theorem for Schrödinger type operators in 2D \cite{Simon:1976un} (see also \cite{Yang}).

Consequently, for type-II strings far from the BPS limit, it becomes the dominant massive mode. In particular, for $\lambda>1$, the vector mode lies below the shape mode until the latter crosses the gauge mass threshold, and hence it may be the more easily excited internal mode in this regime. By contrast, for $\lambda<1$ the hierarchy gets reversed. In the BPS limit, a level crossing between these two modes occurs. This degeneracy is not a coincidence, rather, it is a consequence of the (quantum mechanical) supersymmetric structure arising at $\lambda=1$ \cite{Alonso-Izquierdo:2015tta,Alonso-Izquierdo:2016pcq}. For further details, see Appendix \ref{App.1} .

It is also worth noting that, as already seen in Figure \ref{fig:spectrum}, for $\lambda \to 0$, new vector modes descend from the continuum. In Appendix \ref{App.2}, we show that for sufficiently small $\lambda$, there is an arbitrarily large number of such modes.

\begin{figure}[t]   
    \centering
    \includegraphics[width=1.0\textwidth]{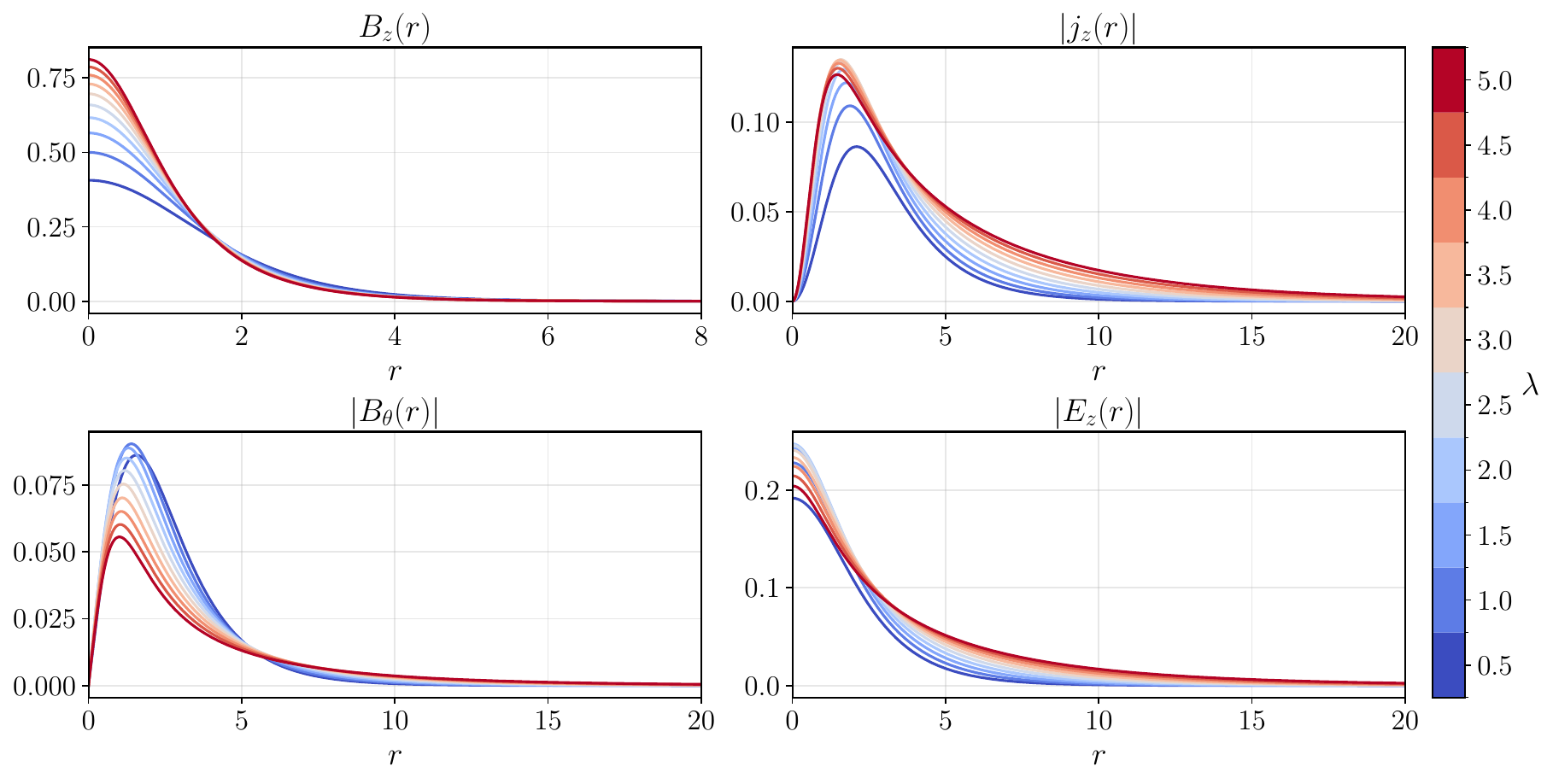}  
    \caption{\justifying
    Radial profiles of the fields and current density for different values of $\lambda$. Upper left: longitudinal magnetic field created by the background vortex string. Upper right: magnitude of the electric current density along $z$ generated by vector mode. Lower left: magnitude of the associated angular magnetic field. Lower right: magnitude of the associated longitudinal electric field.}
    \label{fig: EM fields}
\end{figure}

Just as the zero mode rigidly translates the vortex string in transversal directions, and the shape mode induces radial oscillations of the core, hence modifying its width, the vector mode also has a particular effect on the background soliton. Specifically, its excitation generates both an oscillating electric field along the string, $E_z(r,t)=\partial_tA_z(r,t)$, as well as an oscillating angular magnetic field component, $B_\theta(r,t)=-\partial_rA_z(r,t)$, both illustrated in Figure \ref{fig: EM fields}. 

\begin{figure}[b]   
    \centering
    \includegraphics[width=1.0\textwidth]{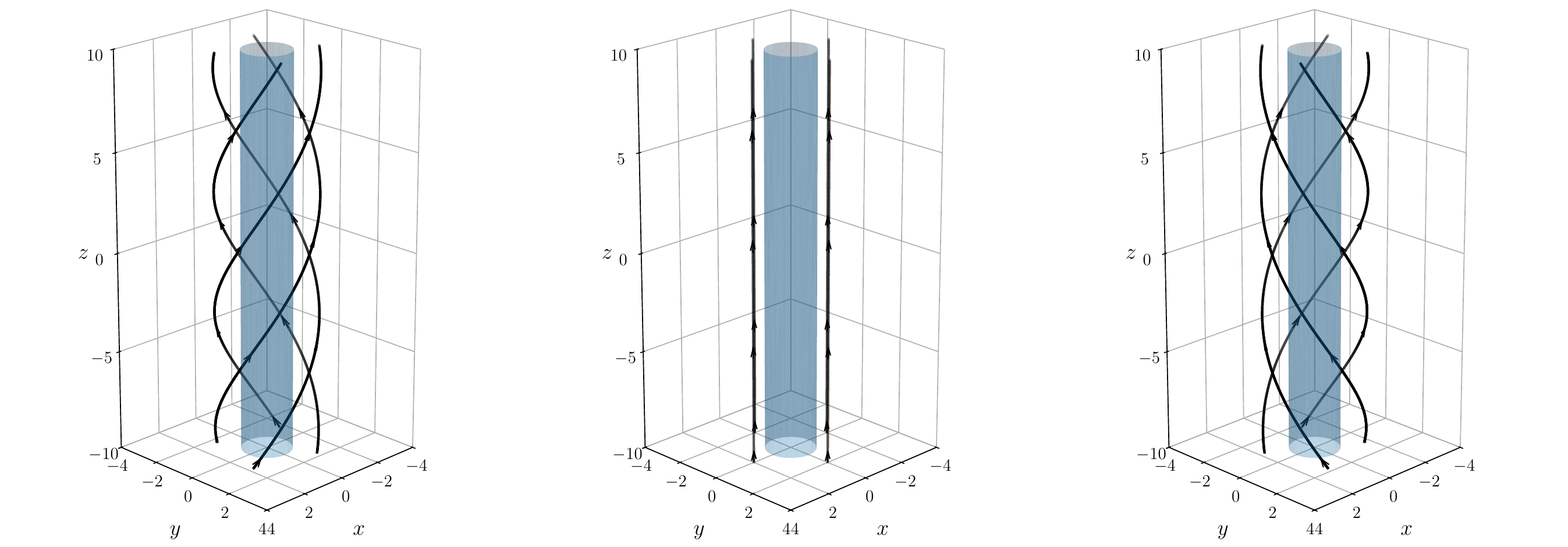}  
    \caption{\justifying
    Pictorial representation of the magnetic field lines of the excited string (blue tube). Between the opposite configurations (left, right), the field lines smoothly transition through the central panel configuration.}
    \label{fig: Magnetic field lines}
\end{figure}

As a result, the latter modifies the otherwise straight magnetic field of the string background, causing its field lines to adopt an helicoidal structure which oscillates periodically (and smoothly) between clockwise and anticlockwise configurations, as seen in \cref{fig: Magnetic field lines}.

\subsection{Physical Picture }

As demonstrated above, an Abelian–Higgs vortex can support localized gauge field excitations that propagate along the string while remaining confined in the transverse directions. From the effective worldsheet perspective, such an excitation behaves as a massive field propagating in one spatial dimension. Its localization originates from the nontrivial radial dependence of the Higgs condensate across the vortex.

The Higgs field is strongly suppressed within the vortex core and vanishes on the string axis. Consequently, the effective mass acquired by the gauge field through the Higgs mechanism is substantially reduced in this region. Far from the vortex, the scalar field approaches its vacuum expectation value and the gauge field recovers its bulk mass. The vortex therefore defines an approximately cylindrical region in which gauge field excitations are energetically favoured. 

An idealized realization of this picture for a straight string is provided by a cylindrical cavity embedded in a superconducting medium. Such a cavity supports purely electromagnetic modes that propagate along its symmetry axis while remaining confined in the transverse plane due to the boundary conditions imposed by the superconduting nature of the surrounding medium. The analogy can be made more quantitative by approximating the system as two distinct regions: an interior in which the gauge field is effectively massless and an exterior in which it possesses a finite mass. Although this sharp interface description neglects the smooth spatial variation of the vortex background, it reproduces several qualitative features of the full fluctuation problem.

Within this approximation, the localized vector excitation may be identified with the lowest axisymmetric transverse magnetic mode of the auxiliary cylindrical waveguide, conventionally denoted by ($\text{TM}_{0 1}$) in pure electromagnetism \cite{Jackson:1998nia}. The field configuration associated with the vortex mode closely parallels that of these electromagnetic waveguide excitation. In particular, it contains an azimuthal magnetic field that is screened by currents localized near the boundary of the effective cavity, resulting in a magnetic field that vanishes asymptotically. The time dependence of this magnetic field is accompanied by a longitudinal electric field, thereby allowing the excitation to propagate along the vortex axis. We see in Fig. \ref{fig: EM fields} that indeed these are the type of fields that represent our localized vector mode in the Abelian-Higgs vortex.

The waveguide interpretation should not, however, be understood as implying that the vortex is a cavity with a sharply defined boundary. The Higgs condensate varies continuously with the radial coordinate, and the exterior therefore constitutes a finite rather than an infinite mass barrier. As a result, the vector mode profile is not restricted to the region conventionally identified as the scalar core, but generally penetrates into the surrounding Higgs phase. This leakage becomes particularly significant when the eigenfrequency approaches the bulk vector mass threshold. In this regime, the mode is only weakly bound and its transverse extent may become larger than the scalar core radius.

The structure of the localized spectrum depends sensitively on the ratio between the scalar and vector mass scales. In the Type I regime, low $\lambda$, the scalar field varies over a comparatively large radial distance, producing a broad region in which the effective gauge field mass is suppressed. The vortex then resembles a wide waveguide capable of supporting several radial standing wave configurations. The lowest modes are comparatively light, while additional localized excitations with one or more radial nodes may also be present. Note that one of such states is already visible in Fig.~\ref{fig:spectrum}. This regime is discussed in detail in Appendix~\ref{App.2}.

As the scalar self coupling is increased, the scalar core becomes progressively narrower. The higher radial excitations, which possess larger transverse gradients and therefore incur a greater gradient energy cost, approach the bulk continuum threshold in succession. Their asymptotic tails become increasingly extended as the threshold is approached. At some point the corresponding profiles cease to be normalizable, and the discrete states become part of the continuum of bulk vector excitations. Modes with the largest number of radial nodes disappear first, leaving only the lowest radial state.

In the Type II regime (high $\lambda$), the scalar core is narrow compared with the scale associated with the vector field. A naive hard wall cavity approximation would then suggest that no bound vector excitation can survive, since confining a mode within such a small region would require a prohibitively large transverse momentum. This conclusion does not apply to the physical vortex, however, because the exterior mass barrier is finite. The lowest mode is not confined entirely within the scalar core, but instead develops a broad tail extending deeply into the surrounding Higgs phase. It consequently becomes a shallow bound state whose effective world sheet mass lies only slightly below the bulk vector mass. Although its binding becomes progressively weaker as the scalar core contracts, the mode persists for every finite value of the coupling parameter.

The persistence of the lowest mode is ultimately related to the spectral properties of attractive potentials in two transverse dimensions. A localized attractive potential, however weak, generically supports at least one bound state under the relevant regularity conditions. For the longitudinal vector perturbation, the local suppression of the Higgs condensate produces precisely such an attractive potential. The lowest vector mode therefore need not disappear at any finite coupling. Instead, its eigenvalue approaches the continuum threshold asymptotically, while its radial profile becomes progressively broader and its binding increasingly weak.

This physical picture and the waveguide analogy applies most directly to the longitudinal vector sector, for which the fluctuation equation reduces to a single channel attractive potential problem. Other gauge field perturbations mix with scalar fluctuations such as the shape mode. Their spectral properties are therefore governed by coupled fluctuation operators and need not obey the same simple arguments based on the electromagnetic field in a cavity intuition. Nevertheless, the waveguide analogy provides a coherent physical description of the longitudinal vector spectrum: a broad Type I vortex can support several light localized modes; increasing the scalar self coupling successively drives the higher radial excitations into the bulk continuum; and a single shallow, spatially extended vector mode persists throughout the strongly Type II regime. All these properties are also reflected in the numerical solutions we found in this paper.


\section{Vector mode decay}\label{sec:rad}
In this section, we want to study both analytically and numerically the non-linear decay of the vector mode due to its higher order coupling to radiation modes. With that purpose in mind, let us extend the previous parametrizations of the fields in order to include radiation fields
\begin{align}\label{eq:ansatz:rad}
    &\phi(r,\theta,t)=e^{i\theta}f(r)+e^{i\theta}R_\phi(r,t)\,, \\
    &A_\theta(r,t)=\frac{a(r)}{r}+R_\theta(r,t)\,,\\
    &A_z(r,t)=C(t)\psi(r)+R_z(r,t)\,.
\end{align}

We assume that the radiation has the same symmetry as the excited vortex. By substituting these field configurations into the equations of motion, one gets, at leading order in the amplitude of the perturbation, $\mathcal{O}\left(C(t)^2\right)$
\begin{align}
    -\ddot R_\phi(r,t)+R_\phi''(r,t)+\frac{R_\phi'(r,t)}{r}&-\left(\frac{3}{2}\lambda f(r)^2-\frac{\lambda}{2}+\frac{(1-a(r))^2}{r^2}\right)R_\phi(r,t)+\notag\\ 
   &\qquad +\frac{2f(r)}{r}(1-a(r))R_\theta(r,t)=C(t)^2\psi(r)^2f(r)\,,   \label{eq_rad_phi} \\
    -\ddot R_\theta(r,t)+R_\theta''(r,t)+\frac{R_\theta'(r,t)}{r}&-\left(f(r)^2+\frac{1}{r^2}\right)R_\theta(r,t)+\frac{2f(r)}{r}(1-a(r))R_\phi(r,t)=0\,,\label{eq_rad_theta} \\ 
    -\ddot R_z(r,t)+R_z''(r,t)+\frac{R_z'(r,t)}{r}&-R_z(r,t)f(r)^2=0 \,. \label{eq_rad_z}
\end{align}

One can see that the vector mode acts as a source for scalar radiation at quadratic order. Furthermore, since gauge ($\theta$ component) and scalar radiation are coupled to one another in equations (\ref{eq_rad_phi}) and (\ref{eq_rad_theta}), the vector mode will also source the former. By contrast, the $z$-component of the gauge field receives no source term at this order, nor is it coupled to the rest of the fields. Hence, we do not expect radiation in this channel at dominant order. 

Given that the amplitude of the vector mode can be expressed in its linear approximation as $C(t)=\hat{C}(t)\cos(\omega_zt)$, where $\hat{C}(t)$ is its envelope amplitude and $\hat{C}(0)=C_0$, one immediately sees that the time-dependent part of the source will oscillate at $2\omega_z$. Therefore, the radiation modes can be decomposed as
\begin{align}
    R_\phi(r,t)&=R_\phi(r)e^{2i\omega_zt}\,,\\
    R_\theta(r,t)&=R_\theta(r)e^{2i\omega_zt}\,,
\end{align}
so that the radiation frequency for the scalar and $\theta$ component of the gauge field, at dominant order, will be twice that of the vector mode. Then, the previous system of differential equations becomes
\begin{align}\label{eq:rad1}
-R_\phi''(r)-\frac{R_\phi'(r)}{r}&+\left(\frac{3}{2}\lambda f(r)^2-\frac{\lambda}{2}+\frac{(1-a(r))^2}{r^2}-4\omega_z^2\right)R_\phi(r)-\notag\\ &\qquad\quad-\frac{2f(r)}{r}(1-a(r))R_\theta(r)=-\frac{\hat{C}(t)^2\psi(r)^2f(r)}{2}\,,\\\label{eq:rad2}
-R_\theta''(r)-\frac{R_\theta'(r)}{r}&+\left(f^2+\frac{1}{r^2}-4\omega_z^2\right)R_\theta(r)-\frac{2f(r)}{r}(1-a(r))R_\phi(r)=0\,,
\end{align}

It is a system of coupled, non-homogeneous and linear differential equations. This system has the same structure as that arising in the Abelian-Higgs model in (2+1) dimensions \cite{Alonso-Izquierdo:2024tjc}, with the only difference being the explicit form of the source terms. 

Several comments are in order. If the source frequency ($2\omega_z$) lies below both mass thresholds, i.e. $2\omega_z < 1$ and $2\omega_z < \sqrt{\lambda}$, the fields $R_{\phi,\theta}$ cannot propagate and therefore the mode does not decay (at quadratic order). This situation does not appear to occur, except possibly in the limit $\lambda \to 0$. 
In the opposite case ($2\omega_z > 1$ and $2\omega_z > \sqrt{\lambda}$), both the scalar and vector channels correspond to scattering modes, and the vector mode decays by emitting radiation. 

There remains one final possibility. If $2\omega_z \le \sqrt{\lambda}$ and $2\omega_z > 1$, it is clear that in this case the scalar channel cannot propagate while the vector channel can. This occurs for $\lambda \gtrsim 3.8$. Beyond this value, the vector mode couples to the quasinormal modes of the string and decays by emitting radiation in the vector channel. It is also possible that, for very small values of $\lambda$, $2\omega_z < 1$ while $2\omega_z > \sqrt{\lambda}$. In this case, the vector mode would also couple to quasinormal modes, whose scalar component is now a scattering state.

Owing to the structural similarity in the system of equations, the methodology developed in \cite{Alonso-Izquierdo:2024tjc} could in principle be applied to solve the present system. Nevertheless, given these different regimes just discussed, it can be more efficient to solve the system numerically, with an appropriate choice of boundary conditions.

Regularity of the solutions at the origin imposes $R_\phi(0)=0$ and $R_\theta(0)=0$. For $r\to\infty$, the choice of BCs becomes less trivial, as they should be chosen with the intention of reproducing the correct asymptotic behavior in each regime.

When $2\omega_z > 1$ and $2\omega_z>\sqrt{\lambda}$, radiation in both fields is expected, and the proper BCs at large radius should select purely outgoing waves, that is,
\begin{eqnarray}
    R_\phi(r)&\propto&\frac{e^{-ik_\phi r}}{\sqrt{r}}\Longrightarrow R'_\phi(L)=\left(-ik_\phi-\frac{1}{2L}\right)R_\phi(L), \quad k_\phi=\sqrt{4\omega_z^2-\lambda}\,,\\
    R_\theta(r)&\propto&\frac{e^{-ik_\theta r}}{\sqrt{r}}\Longrightarrow R'_\theta(L)=\left(-ik_\theta-\frac{1}{2L}\right)R_\theta(L), \quad k_\phi=\sqrt{4\omega_z^2-1}\,,
\end{eqnarray}
 where $L$ marks the boundary of the simulation box.

With the boundary conditions specified, the coupled inhomogeneous system can be solved numerically for a wide range of values of $\lambda$.  The asymptotic form of the resulting fields $R_\phi(r)$ and $R_\theta(r)$ in the regime $\lambda<3.8$ (that is, when $R_\phi$ and $R_\theta$ are asymptotic scattering states) can be approximated as
\begin{align}
    R_\phi(r,t)&\sim \hat{C}(t)^2\frac{A_\phi}{\sqrt{r}}\Re[\exp\{-i(k_\phi r-2\omega_zt-\zeta_\phi)\}]\,,\\
    R_\theta(r,t)&\sim \hat{C}(t)^2\frac{A_\theta}{\sqrt{r}}\Re[\exp\{-i(k_\theta r-2\omega_zt-\zeta_\theta)\}]\,,
\end{align}
where $\zeta_\phi$ and $\zeta_\theta$ are phases irrelevant for our purposes, and the amplitudes $A_\phi$ and $A_\theta$ should be obtained numerically from the large-$r$ behavior of the solutions. From there, one can compute the outgoing energy-flux in the radial direction as
\begin{equation} T_{0r}=\dot{R}_\phi\partial_rR_\phi+\dot{R}_\theta\partial_rR_\theta+\dot{R}_\theta\frac{R_\theta}{r}\,.
\end{equation}

The energy flux averaged over one period gives
\begin{equation}
    \langle T_{0r}\rangle=-\hat{C}(t)^4\frac{\omega_z}{r}\left(k_\phi|A_\phi|^2+k_\theta|A_\theta|^2\right)\,,
\end{equation}
and, thus, the power reads
\begin{equation}
    \dot{E}=L_z\int_0^{2\pi} \langle T_{0r}\rangle r d\theta=-L_z 2\pi\omega_z\hat{C}(t)^4\left(k_\phi|A_\phi|^2+k_\theta|A_\theta|^2\right)\,.
\end{equation}

On the other hand, the energy per unit length associated with the string excited by its vector mode is
\begin{equation}
    \frac{E}{L_z}=\mu_{\rm string}+\frac{\hat{C}(t)^2\omega_z^2}{2}\left(2\pi\int_0^\infty\psi^2(r)r\,dr\right)\,.
\end{equation}

Hence
\begin{equation}
    \mu_{\rm vector\,mode}=\frac{\hat{C}(t)^2\omega_z^2}{2}\left(2\pi \int_0^\infty\psi^2(r)r\,dr\right)=N_{\rm vec}\frac{\hat{C}(t)^2\omega_z^2}{2}\,,
\end{equation}
$N_{\rm vec}$ being the normalization factor of the vector mode ($N_{\rm vec}=1$ from now on unless explicitly stated otherwise).

Now, on the grounds of energy conservation,
\begin{equation}
    \frac{\omega_z^2}{2}\frac{d\hat{C}(t)^2}{dt}=-2\pi\omega_z\hat{C}(t)^4\left(k_\phi|A_\phi|^2+k_\theta|A_\theta|^2\right)\,,
\end{equation}
 one can derive a decay law for the vector mode's amplitude of the form:
 \begin{equation}\label{eq: vector_decay}
     \hat{C}(t)=\frac{1}{\sqrt{\Gamma(\lambda) t+\hat{C}(0)^{-2}}}\,, \qquad\Gamma(\lambda)=\frac{4\pi\left(k_\phi|A_\phi|^2+k_\theta|A_\theta|^2\right)}{\omega_z}\,.
 \end{equation}
This power law decay is formally identical to that of the shape mode in the same model.
  
 The regime $\lambda\gtrsim3.8$ requires a separate explanation. As we have mentioned, in this situation we have  $2\omega_z \le \sqrt{\lambda}$ and $2\omega_z > 1$ and the vector mode couples to the quasinormal modes of the string.  Since $R_\phi$ cannot propagate (bound component of the quasinormal mode), the energy of the shape mode can be radiated only through $R_\theta$. 
The BCs for $R_\theta(r)$ should be maintained as before, while those for $R_\phi(r)$ should match an exponentially decaying solution, namely

\begin{equation}
    R_\phi(r)\propto \frac{e^{-\kappa_\phi r}}{\sqrt{r}}\Longrightarrow R'_\phi(L)=\left(-\kappa_\phi-\frac{1}{2 L}\right)R_\phi(L), \quad \kappa_\phi=\sqrt{\lambda-4\omega_z^2}\,.
\end{equation}

 Initially, the (bound) scalar component of the mode is excited and oscillates with frequency $2\omega_z$. It couples linearly to the vector component of the mode, which radiates. As we will see, the energy required to initially excite the scalar component of the mode is negligible compared to that of the vector mode, so one can assume that all dissipation occurs through the vector channel. Under this approximation, we can replace (\ref{eq: vector_decay}) by
 \begin{equation}\label{eq:vector_decay_1}
\Gamma(\lambda)\approx\frac{4\pi\left(k_\theta|A_\theta|^2\right)}{\omega_z}\,.
 \end{equation}
 
\begin{figure}[h]   
    \centering
    \includegraphics[width=0.8\textwidth]{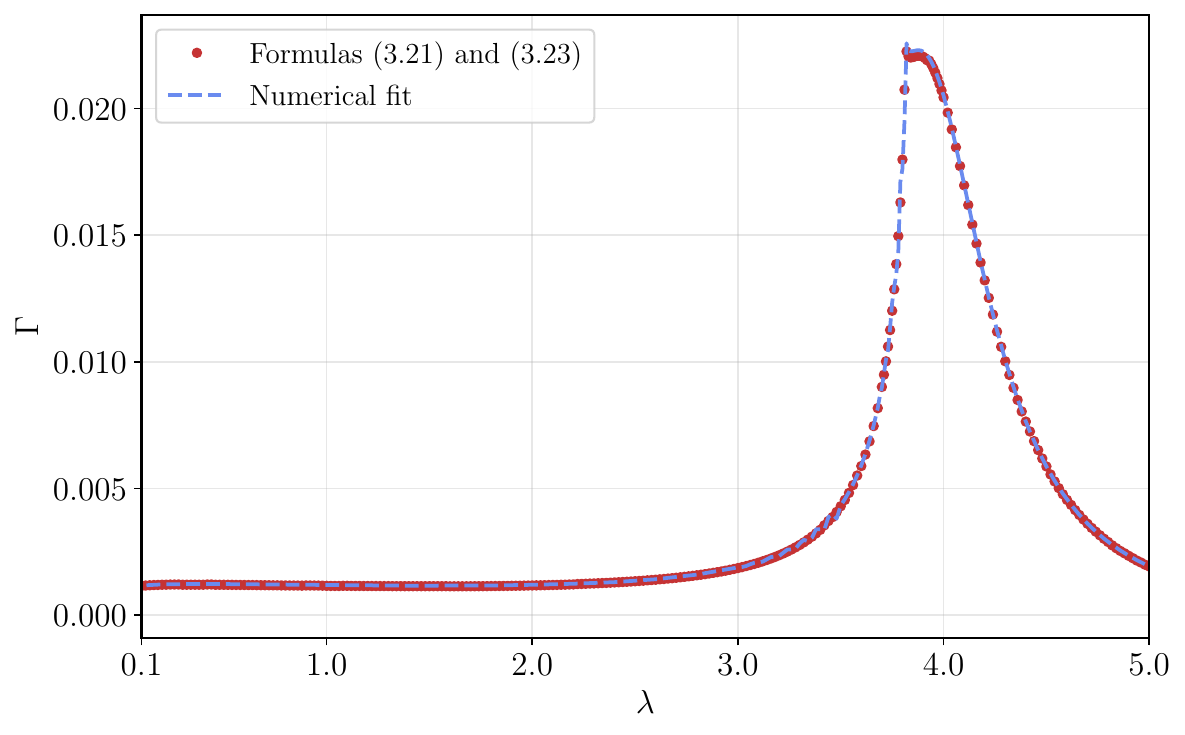}  
    \caption{\justifying
    Decay constant of the vector mode amplitude as a function of the dimensionless coupling constant $\lambda$. The blue discontinuous line shows results from a power law fit to the numerical field theory decay. The red dots correspond to values extracted from the system of inhomogeneous equations (\ref{eq:rad1})–(\ref{eq:rad2}), following formulas (\ref{eq: vector_decay}) and (\ref{eq:vector_decay_1}). For clarity near the region of the peak, in the interval $3.7\leq\lambda\leq4.0$ we use a spacing $\Delta\lambda=0.01$, while $\Delta\lambda=0.02$ is used for the rest of the displayed range.}
    \label{fig: decay_const_vs_coupling}
\end{figure}

 As an \textit{a posteriori} justification of this approximation, we can see in Fig. \ref{fig: decay_const_vs_coupling} that the analytical prediction agrees with our numerics. However, situations in which the scalar component can be excited with a non-negligible fraction of the available excitation energy require a more refined treatment.

The behavior of the decay constant shown in \cref{fig: decay_const_vs_coupling} can be mainly understood by means of the overlap between the source and the (scalar) radiation modes. In the $plateau$ region, the radiation modes are highly oscillatory, which suppresses their overlap with the localized source profile. As the coupling grows and the frequency of the scalar radiation approaches the scalar mass threshold from above, its wavelength grows and so does its overlap with the source, as they become less oscillatory. As a consequence, the decay constant rises gradually until it peaks at around $\lambda\sim3.81$. From that point onward, the frequency of the scalar radiation falls below the mass threshold (See \cref{fig:spectrum}), and the modes become non-propagating. Even if in this regime the overlap with the source is still considerable, radiation can only leak away through the vector channel, which is sourced indirectly through its coupling to the scalar field. Therefore, the energy leakage becomes less efficient, and the decay constant decreases. We note that for higher values of $\lambda$ not present in the figure, we have not observed any deviation from the expected decrease in the decay constant.

We have tested this semianalytical power law decay's validity against numerical field theory simulations. In these simulations, we have initialized the static string excited with the vector mode, and let it evolve in time, while monitoring its amplitude. The cylindrical symmetry of the problem has allowed us to solve the equations of motion in a 1+1 dimensional setup, in terms of just the radial coordinate. It is worth mentioning that in this numerical scheme, extracting the vector mode amplitude is particularly straightforward, as the vortex string is pinned at the origin and there is no background field for $A_z(r,t)$, but only mutually orthogonal perturbations\footnote{For further details in the numerical scheme, see Appendix \ref{App.3}.}. As a result, the amplitude can be obtained by simply projecting the $z$ component of the gauge field read off from the simulation over the vector mode's profile, that is,
\begin{equation}
    C(t)=\int_0^{2\pi}d\theta\int_0^\infty A_z(r,t)\psi(r) rdr\,.
\end{equation}

\begin{figure}[h]
    \centering
    \includegraphics[width=1.0\textwidth]{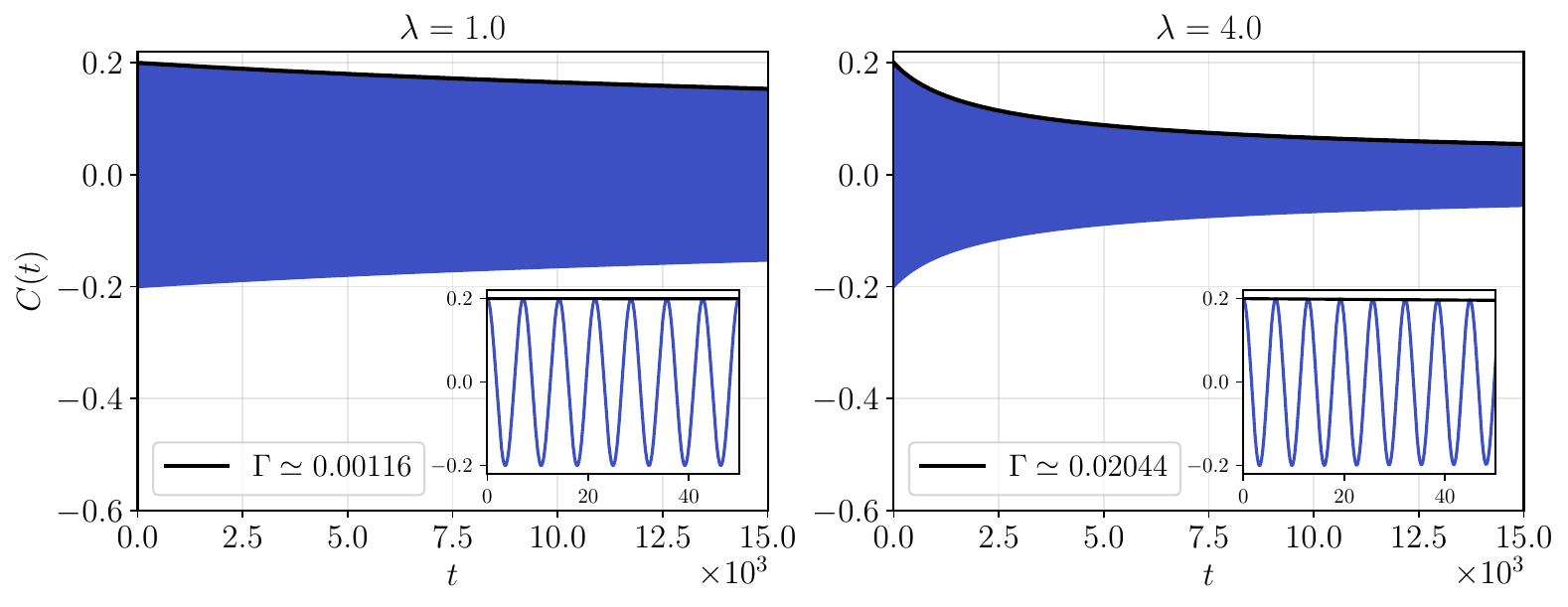} 
    \caption{\justifying
    Amplitude of the vector mode, $C(t)$, as a function of time extracted from the field theory simulations for $\lambda=1.0$ (left) and $\lambda=4.0$ (right). The black curves correspond to the analytical decay, with $\Gamma$ being the decay constant of the power law decay formula (\ref{eq: vector_decay}). The zoomed plots display the first oscillations of the amplitude.}
    \label{fig: VecModeDecayComparison}
\end{figure}

Fig. \ref{fig: VecModeDecayComparison} shows a direct comparison between the numerically extracted amplitude's decay and the analytically predicted one via \cref{eq: vector_decay} for two distinct coupling constants, each corresponding to one of the two regimes just discussed. The analytical prediction, shown in the plots by a black curve, seems to be in perfect agreement with the numerical findings in both cases. As mentioned previously, for values of the couplings close to where $2\omega_z$ crosses the scalar threshold, the initial decay is more pronounced. Overall, the vector mode appears to be as effective as the shape modes at storing energy for long time scales.

\begin{figure}[h]   
    \centering
    \includegraphics[width=1.0\textwidth]{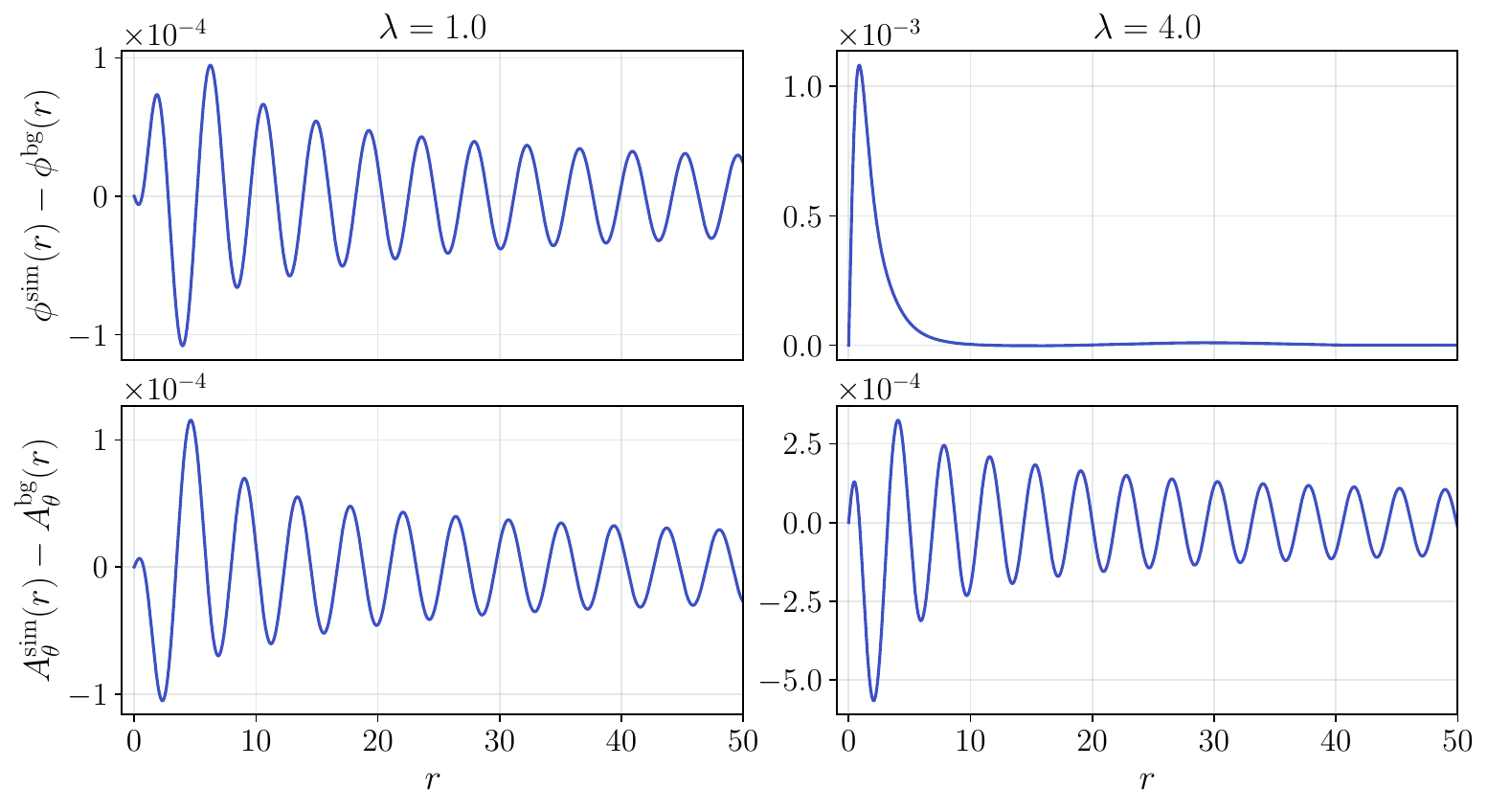}  
    \caption{\justifying
    Snapshots of the radial profiles of the scalar and gauge field perturbations extracted from the simulation at $t=905.0$, for two different values of $\lambda$. The profiles are obtained by subtracting the background vortex profiles from the corresponding evolved fields.}
    \label{fig: radiation fields}
\end{figure}

Regarding the emitted radiation, as expected, we encounter fundamental differences between both cases. In Fig. \ref{fig: radiation fields} we show a snapshot of the scalar and angular gauge perturbation field profiles from the previous decays. In accordance with what we argued before, for the BPS case both channels radiate equally efficiently. By contrast, for $\lambda=4.0$ only radiation in the gauge channel is appreciated at dominant, quadratic order, while the scalar profile shows a ``lump-like" configuration, exponentially localized close to the string core. In other words, for $\lambda\gtrsim 3.81$, the decay of the vector mode leads to the excitation of one of the multiple Feshbach resonances encountered in the spectrum of the local string for this parameter region \cite{Alonso-Izquierdo:2024tjc}.

On the other hand, the Fourier analysis shows that radiation in these two channels, whenever they are open, is dominated by a frequency twice that of the vector mode, as anticipated at the beginning of the section. As far as the $z$ gauge radiation is concerned, we have seen that the amplitude of the radiation tail is orders of magnitude lower than that of the scalar and angular channels, while its frequencies are multiples of $\omega_z$, corresponding to higher order terms in the perturbative expansion.


\section{Interaction and instability: vector vs. zero mode}
\label{Sec: interac}

Parametric instabilities are common phenomena, perhaps universal, in the dynamics of several types of (excited) strings  \cite{Blanco-Pillado:2022rad,Blanco-Pillado:2022axf,Aurrekoetxea:2026xrl}, where a resonant energy transfer takes place between a massive internal mode of the string and the Goldstone sector (i.e. a transversal zero mode). This feature accelerates the radiation emission by the massive bound state under study, which are typically shape modes. In this section, we show that, similarly to the shape mode, a parametric instability can also be triggered between the vector and zero modes, albeit with some notable differences.

A standard way to study this kind of phenomenon, complementary to field theory simulations, is through the so-called collective coordinate model (CCM). In this approach, the field theory Lagrangian, which contains infinitely many degrees of freedom, is reduced to an effective mechanical one, with a finite number of them. This truncation is implemented by choosing an appropriate ansatz for the fields. In general, it will consist of the static solitonic solution plus a finite set of internal modes, whose time dependent amplitudes will be the collective coordinates of the model.

This approach can not only reproduce the field theoretical results to a good extent but, most importantly, it also provides insight into the system under study by identifying which modes (or interactions between modes) are responsible for the observed nonlinear dynamics. Of course, the success of the CCM will heavily depend on the chosen ansatz. If relevant modes are omitted, it can lead to incorrect resonance conditions between mode frequencies or fail to accurately reproduce the time evolution of the amplitudes seen in the full field theory simulation.

In previous studies, the essential dynamics of the parametric instability has been captured by an effective model containing only two dynamical amplitudes: the one of the massive internal mode and the amplitude of one zero mode. By direct analogy, it is natural to ask whether an instability involving the vector mode can be described by a similar model with the same number of variables. We therefore consider an ansatz composed by a straight string with a uniformly excited vector mode and a single zero mode in the $x$ direction, that is,
\begin{align}
    \phi(r,\theta,t)&=e^{i\theta}f(r)+X(t)\delta_x\phi^v(r,\theta)\cos(k_0z)\,,\\
    A_r(r,\theta,t)&=X(t)\delta_xA_r^v(r,\theta)\cos(k_0z)\,,\\
    A_\theta(r,\theta,t)&=\frac{a(r)}{r}+X(t)\delta_xA_\theta^v(r,\theta)\cos(k_0z)\,,\\
    A_z(r,t)&=C(t)\psi(r),
\end{align}
where $X(t)$ is the amplitude of the zero mode, and the zero mode functions read
\begin{align}
\delta_x\phi^v&=D_x\phi=e^{i \theta}\left[\cos\theta\partial_r+i\frac{\sin\theta}{r}(a-1)\right]f\,,\\
    \delta_xA_r^v&=\frac{a'}{r}\sin\theta\,,\\
    \delta_xA_\theta^v&=\frac{a'}{r}\cos\theta\,.
\end{align}

 It can be checked that these fluctuations fulfill the background gauge condition \cite{Goodband:1995rt, Cheng:1984vwu}, so their effect on the background soliton cannot be gauged away. Substituting this ansatz into the original Lagrangian (\ref{dimless_lag}) and integrating along spatial directions yields the following {\it mechanical Lagrangian},
\begin{equation}
    L_{\rm{eff}}=I_0+I_{dX2}\dot X(t)^2+I_{dC2}\dot C^2(t)+I_{X2}X(t)^2+I_{C2}C^2(t)+I_{X4}X(t)^4+I_{X2C2}X(t)^2C(t)^2\,,\label{eq: EffLag2DoF}
\end{equation}
where each of the coefficients multiplying the amplitudes corresponds to the aforementioned integrals. Their expressions can be found in Appendix \ref{App.4}. The equations of motion for the mode amplitudes are
\begin{align}
    \ddot C(t)-\left(\frac{I_{C2}}{I_{dC2}}+\frac{I_{X2C2}}{I_{dC2}}X(t)^2\right)C(t)&=0\,,\\
    \ddot X(t)-\left(\frac{I_{X2}}{I_{dX2}}+2\frac{I_{X4}}{I_{dX2}}X(t)^2+\frac{I_{X2C2}}{I_{dX2}}C(t)^2\right)X(t)&=0\,,
\end{align}
with
\begin{equation}
    \frac{I_{X2}}{I_{dX2}}=-k_0^2\,,\quad \frac{I_{C2}}{I_{dC2}}=-\omega_z^2\,.
\end{equation}

Let us consider the equation for the zero mode amplitude first. At lowest non-linear order, the equation has the following form
\begin{equation}
    \ddot X(t)+\left(k_0^2-\frac{I_{X2C2}}{I_{dX2}}C^2(t)\right)X(t)=0~.
\end{equation}

Assuming that the vector mode amplitude is given by its linear approximation, i.e., $C(t)=C_0\cos(\omega_z t)$, we obtain the following Mathieu equation for the zero mode amplitude,
\begin{equation}
    \ddot X(t)+\left[\left(k_0^2-\frac{C_0^2}{2}\frac{I_{X2C2}}{I_{dX2}}\right)-\frac{C_0^2}{2}\frac{I_{X2C2}}{I_{dX2}}\cos(2\omega_z t)\right]X(t)=0\,,
\end{equation}
whose principal resonance condition, for small enough amplitudes of the vector mode, can be shown to be
\begin{equation}
    k_0=\sqrt{\omega_z^2+\frac{C_0^2}{2}\frac{I_{X2C2}}{I_{dX2}}}\approx \omega_z\,.
\end{equation}

Likewise, an identical computation gives a similar resonance condition for the equation of the vector mode amplitude, namely
\begin{equation}
    \omega_z=\sqrt{k_0^2+\frac{C_0^2}{2}\frac{I_{X2C2}}{I_{dC2}}}\,.
\end{equation}

This prediction on the resonance condition can be verified in the full field theory by initializing the string with a uniform vector mode excitation and a smaller translational mode, while choosing the size of the box along which the string ($z$ direction in our case) so that $k_0\approx\omega_z$ is satisfied.

\begin{figure}[h]   
    \centering
    \includegraphics[width=0.8\textwidth]{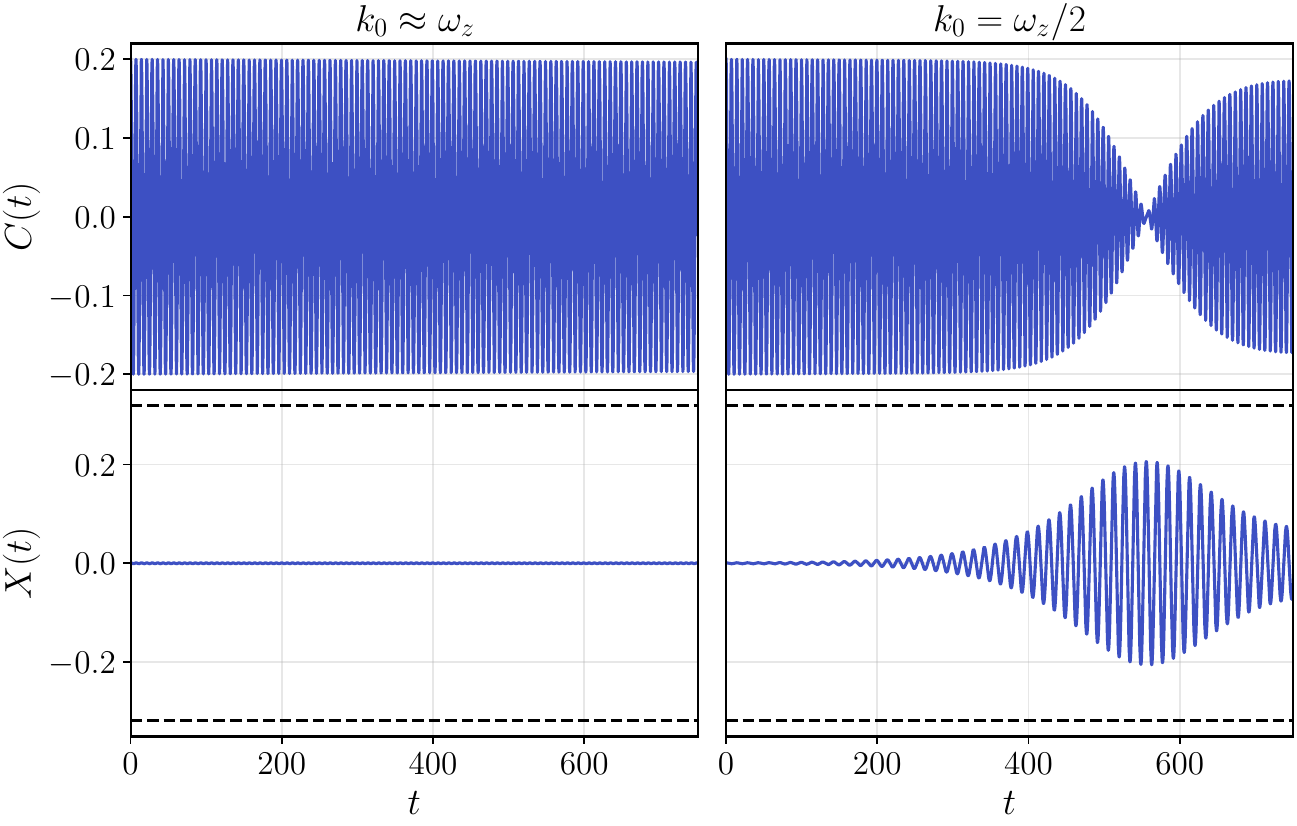}  
    
    \caption{\justifying
    Evolution of the extracted vector and zero mode amplitudes in the BPS case for two different choices of the wavenumber $k_0$. The length of the simulation box along the direction where the string lies has been chosen to be $L_z=14.25$, such that both wavenumbers $\left(k_0=2\pi/L_z\simeq\omega_z/2\right)$ and $\left(k_0=4\pi/L_z\simeq\omega_z\right)$ fit, given that for $\lambda=1$, $\omega_z\sim0.882$. The initial amplitudes have been chosen to be $C_0=0.2$ and $X_0=10^{-3}$. The horizontal dashed lines in the lower graphs show the analytical estimate for the maximum amplitude of the zero mode.}
    \label{fig: resonance_comparison}
\end{figure}

However, as seen in Fig. \ref{fig: resonance_comparison}, the full field theory simulation does not show the parametric resonance predicted by this minimal truncation. Instead, a parametric growth in the zero mode amplitude is observed when $k_0=\omega_z/2$ is fulfilled. This discrepancy between the predicted condition in the CCM and the results from the field theory simulation seems to indicate that the ansatz containing just two degrees of freedom is insufficient to explain the observed dynamics and should be extended by considering at least another mode. Indeed, a resonance occurring at $k_0=\omega_z/2$ could be directly explained by the presence of a quadratic term linear in $C(t)$ at the level of the equation of motion for the zero mode amplitude.

Furthermore, by energy conservation arguments, we can infer the maximum amplitude that the zero mode can reach once it is reasonantly amplified. At leading order, the energy per unit length of the vector mode and zero mode can be respectively written as $E_{\rm{vec}}=N_{\rm vec}C_0^2\omega_z^2/2$ and $E_{\rm{X}}=N_{\rm X}X_0^2k_0^2/2$, with $N_{\rm vec}$ and $N_X$ being the normalization factors. For a configuration with initial conditions as the one in Fig. \ref{fig: resonance_comparison}, assuming that all the energy is cleanly transferred from the vector to the zero mode, its maximum amplitude can be estimated to be $X(t_{\rm{peak}})=(C_0\omega_z/k_0)\sqrt{N_C/N_X}\approx0.319$. Nevertheless, even in the resonant case, we can see that the maximum amplitude attained by the zero mode remains well below the estimated value. This is consistent with part of the extra energy stored in the vector mode being transferred to additional internal degrees of freedom, and reinforces the idea of going beyond the current CCM \footnote{In principle, this argument alone can also point towards the rest of the extra energy being entirely lost in the form of radiation. However, as we shall show next, this is not the case.}.

\begin{figure}[h]   
    \centering
    \includegraphics[width=\textwidth]{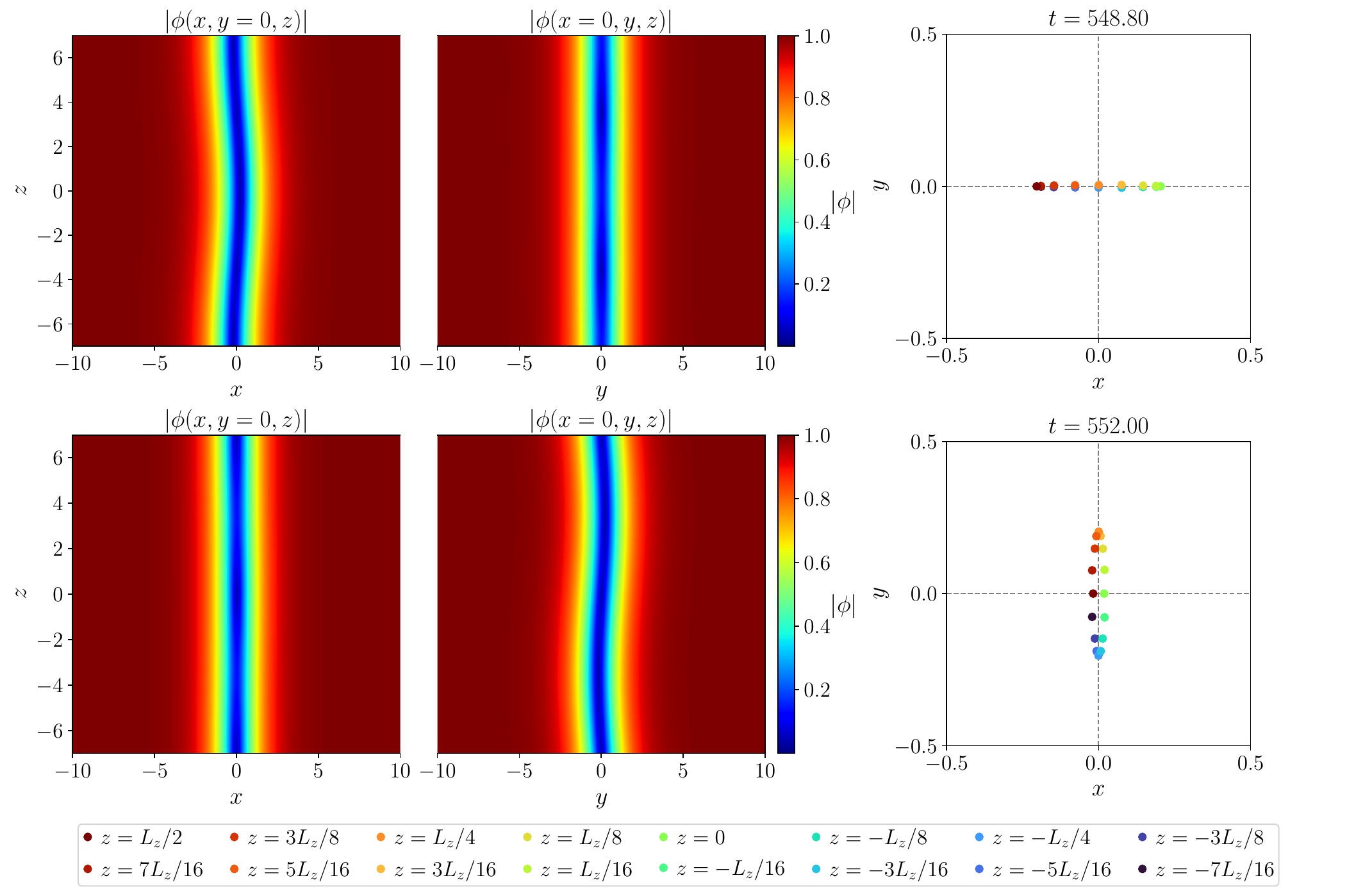}  
    
    \caption{\justifying
    Snapshots of the string configuration during the parametric resonance. The left and middle panels show the modulus of the scalar field in the $x - z$ and $y - z$ planes, respectively, while the right panels display the reconstructed string core projected onto the $x - y$ plane, with each point corresponding to a different value of $z$. Additional snapshots of the evolution are provided in Appendix~\ref{App:5}, while the corresponding videos can be found in Ref.~\cite{ancillary_videos}.}
    \label{fig: snapshots_first_case}
\end{figure}

Fig. \ref{fig: snapshots_first_case} provides further evidence that the current effective model is too restrictive by identifying one of the omitted degrees of freedom. The left and middle panels show snapshots of the scalar profiles of the string taken at different times. As we can see, close to where the zero mode amplitude $X(t)$ is maximum, at $t=548.8$, the string is clearly excited along the $x$ direction. After approximately a quarter of an oscillation period has elapsed, however, the transverse deformation has rotated towards the $y$ direction.  This behavior is particularly evident from the trajectories of the string centers displayed in the right panels. The full field theory evolution therefore demonstrates that excitation of the vector mode induces motion in both transverse components of the translational zero mode. The effective model must consequently be extended to include the translational degree of freedom associated with displacements along the $y$ direction. In addition, for reasons that will become clear below, the scalar shape mode must also be incorporated. We now construct the corresponding extended effective model.

The field configurations are now
\begin{align}
    \phi(r,\theta,z,t)&=e^{i\theta}f(r)+X(t)\delta_x\phi^v\cos(k_0z)+Y(t)\delta_y\phi^v\cos(k_0z+\delta)+S(t)\varphi(r)e^{i\theta}\cos(k_sz)\,,\\
    A_r(r,\theta,z,t)&=X(t)\delta_xA_r^v\cos(k_0z)+Y(t)\delta_yA_r^v\cos(k_0z+\delta)\,,\\
    A_\theta(r,\theta,z,t)&=\frac{a(r)}{r}+X(t)\delta_xA_\theta^v\cos(k_0z)+Y(t)\delta_yA_\theta^v\cos(k_0z+\delta)+S(t)\alpha(r)\cos(k_sz)\,,\\
    A_z(r,t)&=C(t)\psi(r)\,,
\end{align}
where $Y(t)$ and $S(t)$ respectively represent the amplitudes of the zero mode in the $y$ direction and the shape mode, and $\delta$ is the relative phase shift between the modulations of the two zero modes along $z$. The explicit expressions for the zero mode in the $y$ direction are 
\begin{align}
    \delta_y\phi^v&=D_y\phi=e^{i \theta}\left[\sin\theta\partial_r-i\frac{\cos\theta}{r}(a-1)\right]f,\\
    \delta_yA_r^v&=-\frac{a'}{r}\cos\theta\,,\\
    \delta_yA_\theta^v&=\frac{a'}{r}\sin\theta\,.
\end{align}
After substituting the expressions for the fields in (\ref{dimless_lag}) and integrating along spatial directions, the effective Lagrangian with four degrees of freedom reads
\begin{align}
L=\,&I_{dX2}\dot X(t)^2+I_{dY2}\dot Y(t)^2+I_{dC2}\dot C(t)^2+I_{dS2}\dot S(t)^2+I_{X2}X(t)^2+I_{Y2}Y(t)^2+I_{C2}C(t)^2+I_{S2}S(t)^2\notag\\
&+I_{X4}X(t)^4+I_{Y4}Y(t)^4 +I_{S3}S(t)^3+I_{S4}S(t)^4\notag\\
&+I_{X2S}X(t)^2S(t) +I_{Y2S}Y(t)^2S(t)+I_{XYC}X(t)Y(t)C(t)+I_{C2S}C(t)^2S(t)\notag\\
&+I_{X2Y2}X(t)^2Y(t)^2+I_{X2C2}X(t)^2C(t)^2+I_{Y2C2}Y(t)^2C(t)^2\notag\\
&+I_{X2S2}X(t)^2S(t)^2+I_{Y2S2}Y(t)^2S(t)^2+I_{C2S2}C(t)^2S(t)^2 .\label{eq: EffLag4DoF}
\end{align}

The key point here is that by allowing the zero mode in the $y$ direction to have a nonzero relative phase shift with respect to its perpendicular counterpart, a cubic term of the form $X(t)Y(t)C(t)$ arises in the mechanical Lagrangian \footnote{For convenience, let us mention the explicit form of this integral here:
\begin{equation}
    I_{XYC}=k_0\sin\delta\int dz\int_0^{2\pi}d\theta\int_0^\infty r dr\left(\frac{(a-1)ff'\psi}{r}\right)\,.\notag
\end{equation}}. Varying this interaction term with respect to $Y$ generates a contribution proportional to $C(t)X(t)$, which acts as a source for the $y$ zero mode. At the level of the field theory equations, an analogous contribution can be identified in the scalar EOM through the term $2iA_z\partial_z\phi$. For $A_z=C(t)\psi(r)$ and $\phi=X(t)\delta_x\phi^v(r,\theta)\cos(k_0z)$, the longitudinal derivative changes the modulation of that term to $\sin(k_0z)$. Consequently, the overlap of this term with the zero mode in $y$ is nonzero only if the latter contains a $\sin(k_0z)$ component in their longitudinal modulation. Thus, in the presence of the vector mode, an $x$ zero mode with $\cos(k_0z)$ modulation will excite a zero mode in $y$ proportional to $\sin(k_0z)$. This expectation can be further verified in the snapshots shown in Fig. \ref{fig: snapshots_first_case}. It is therefore natural to adopt this modulation for the $y$ zero mode in what follows. 

The equations of motion are therefore
\begin{align}
&\ddot X-\frac{I_{X2}}{I_{dX2}}X-\frac{I_{X2S}}{I_{dX2}}XS-\frac{2I_{X4}}{I_{dX2}}X^3-\frac{I_{X2Y2}}{I_{dX2}}XY^2-\frac{I_{X2C2}}{I_{dX2}}XC^2-\frac{I_{X2S2}}{I_{dX2}}XS^2=\frac{I_{XYC}}{2I_{dX2}}YC \,,\label{4DoF_EOM_X}\\
&\ddot Y-\frac{I_{Y2}}{I_{dY2}}Y-\frac{I_{Y2S}}{I_{dY2}}YS-\frac{2I_{Y4}}{I_{dY2}}Y^3-\frac{I_{X2Y2}}{I_{dY2}}X^2Y-\frac{I_{Y2C2}}{I_{dY2}}YC^2-\frac{I_{Y2S2}}{I_{dY2}}YS^2
=\frac{I_{XYC}}{2I_{dY2}}XC \,,\label{4DoF_EOM_Y}\\
&\ddot C-\frac{I_{C2}}{I_{dC2}}C-\frac{I_{C2S}}{I_{dC2}}CS-\frac{I_{X2C2}}{I_{dC2}}X^2C-\frac{I_{Y2C2}}{I_{dC2}}Y^2C-\frac{I_{C2S2}}{I_{dC2}}CS^2=\frac{I_{XYC}}{2I_{dC2}}XY \,,\label{4DoF_EOM_C}\\
&\ddot S-\frac{I_{S2}}{I_{dS2}}S-\frac{3I_{S3}}{2I_{dS2}}S^2-\frac{2I_{S4}}{I_{dS2}}S^3-\frac{I_{X2S2}}{I_{dS2}}X^2S-\frac{I_{Y2S2}}{I_{dS2}}Y^2S-\frac{I_{C2S2}}{I_{dS2}}C^2S=\notag\\
&\hspace{8cm}=\frac{I_{X2S}}{2I_{dS2}}X^2
+\frac{I_{Y2S}}{2I_{dS2}}Y^2
+\frac{I_{C2S}}{2I_{dS2}}C^2 \,.\label{4DoF_EOM_S}
\end{align}

It can be seen that the addition of these 2 extra degrees of freedom has drastically changed the nature of the effective EOMs. Most importantly, higher order terms aside, the vector mode mediates a ``cross coupling" between the two zero modes: $X(t)$ is sourced by $Y(t)C(t)$ and $Y(t)$ by $X(t)C(t)$. On the other hand, the shape mode is sourced by the rest of the excitations.

Hence, the disagreement between the parametric instability condition predicted by the previous CCM and the one observed in the field theory simulations can be understood as follows. Assuming an initial hierarchy for the mode amplitudes as in the previous case, while maintaining the newly added perturbations initially unexcited, the equations for the zero mode amplitudes read, at lowest non-linear order
\begin{equation}\label{eq: XYCeqs}
    \ddot X+k_0^2X=\frac{I_{XYC}}{2I_{dX2}}YC\,,\qquad\ddot Y+k_0^2Y=\frac{I_{XYC}}{2I_{dY2}}XC\,,
\end{equation}
with $I_{dX2}=I_{dY2}$. Thus, an initially unexcited zero mode $Y$ can be sourced due to the presence of its perpendicular counterpart $X$ and the vector mode. Once excited, $Y$ will become a source of $X$, and the cycle will repeat again, gradually increasing their amplitudes until higher order terms become important. This mechanism can be further enhanced if the resonance condition is met. To see this, we can convert the previous system into two Mathieu equations by defining $B_\pm=X\pm Y$ and taking the linearized approximation for the vector mode's amplitude, $C(t)=C_0\cos(\omega_zt)$, such that
\begin{equation}
    \ddot B_\pm(t)+\left(k_0^2\mp C_0\frac{I_{XYC}}{2I_{dX2}}\cos(\omega_zt)\right)B_\pm(t)=0.
\end{equation}

\begin{figure}[h]   
    \centering
    \includegraphics[width=1.0\textwidth]{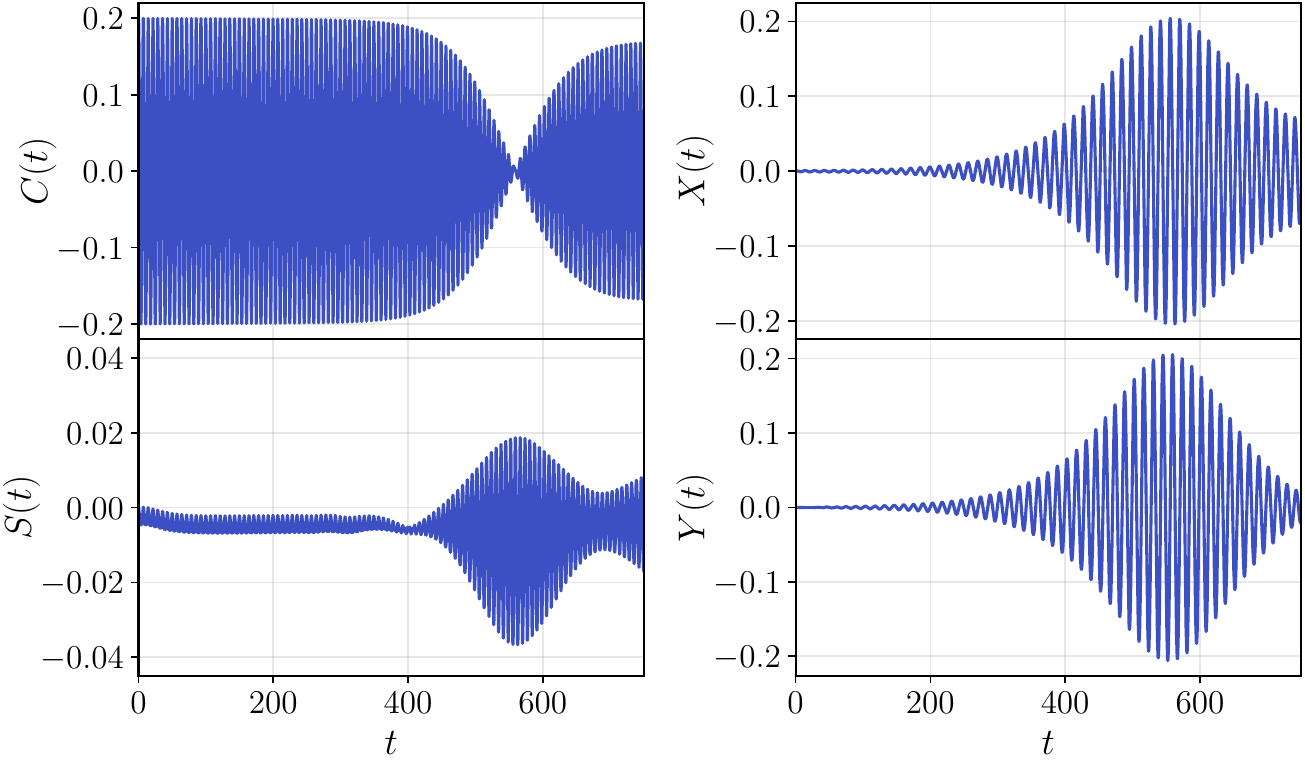}  
    \caption{\justifying
    Time evolution of the amplitudes of the vector, shape and zero modes, extracted from the field theory simulation.}
    \label{fig: resonant amplitudes}
\end{figure}

In this case, the principal instability will always be present at $k_0=\omega_z/2$. In Fig. \ref{fig: resonant amplitudes} we show the time evolution of the four different amplitudes considered in the extended CCM. It demonstrates how, as predicted through this effective model, the zero mode in the $y$ direction resonates exactly as its counterpart, albeit with a small temporal shift, even if its initial amplitude is null. Furthermore, we can see that a smaller amount of energy is also transmitted to the shape mode, which is in accordance with the fact that the rest of the modes source it in the effective EOM (\ref{4DoF_EOM_S}). Thus, we can conclude that the addition of the extra 2 DoF solves the  disagreement between the initial CCM and the observed field theory dynamics.

This effective model provides a much wider range of predictions which can be tested via field theory simulations. For instance, we may slightly change the initial conditions of the previous case by giving the zero mode in $y$ the same amplitude as the one in $x$. By inspecting the EOMs, we expect that, similarly to the previous case, both zero mode amplitudes should become resonantly excited with frequency $k_0=\omega_z/2$. Since now their initial amplitudes are equal, they should evolve in phase with one another, thereby affecting the shape mode amplitude's evolution.

At lowest nonlinear order, the equation for the homogeneous shape mode amplitude reads 
\begin{equation}\label{eq: }
    \ddot S+\omega_s^2S=\frac{I_{X2S}X_0^2+I_{Y2S}Y_0^2}{4I_{dS2}}\left(1+\cos(2k_0t)\right)+\frac{I_{C2S}}{4I_{dS2}}C_0^2\left(1+\cos(2\omega_zt)\right) \,.
\end{equation}

This driven harmonic oscillator equation presents two possible resonance conditions: $\omega_s=2\omega_z$ and $\omega_s=2 k_0$. For the current initial conditions, the latter one is met. We therefore expect the shape mode to undergo driven resonant growth \footnote{This contrasts with the previous case, where $Y(0)=0$. There, at early times, $X(t)\simeq X_0\cos(k_0t)$ and $C(t)\simeq C_0\cos(2k_0t)$, so by looking at equation (\ref{eq: XYCeqs}), we can see that $X(t)C(t)$ sources $Y(t)\propto t\sin(k_0t)$. The two zero modes become approximately shifted in time by a quarter of a period, and therefore their contributions to the oscillatory driving term in the shape mode equation get partially canceled.}. Consequently, the time evolution should exhibit a chain of resonances of the form vector$\,\xrightarrow{~}\,$zero$\,\xrightarrow{~}\,$shape.

\begin{figure}[h]   
    \centering
    \includegraphics[width=0.9\textwidth]{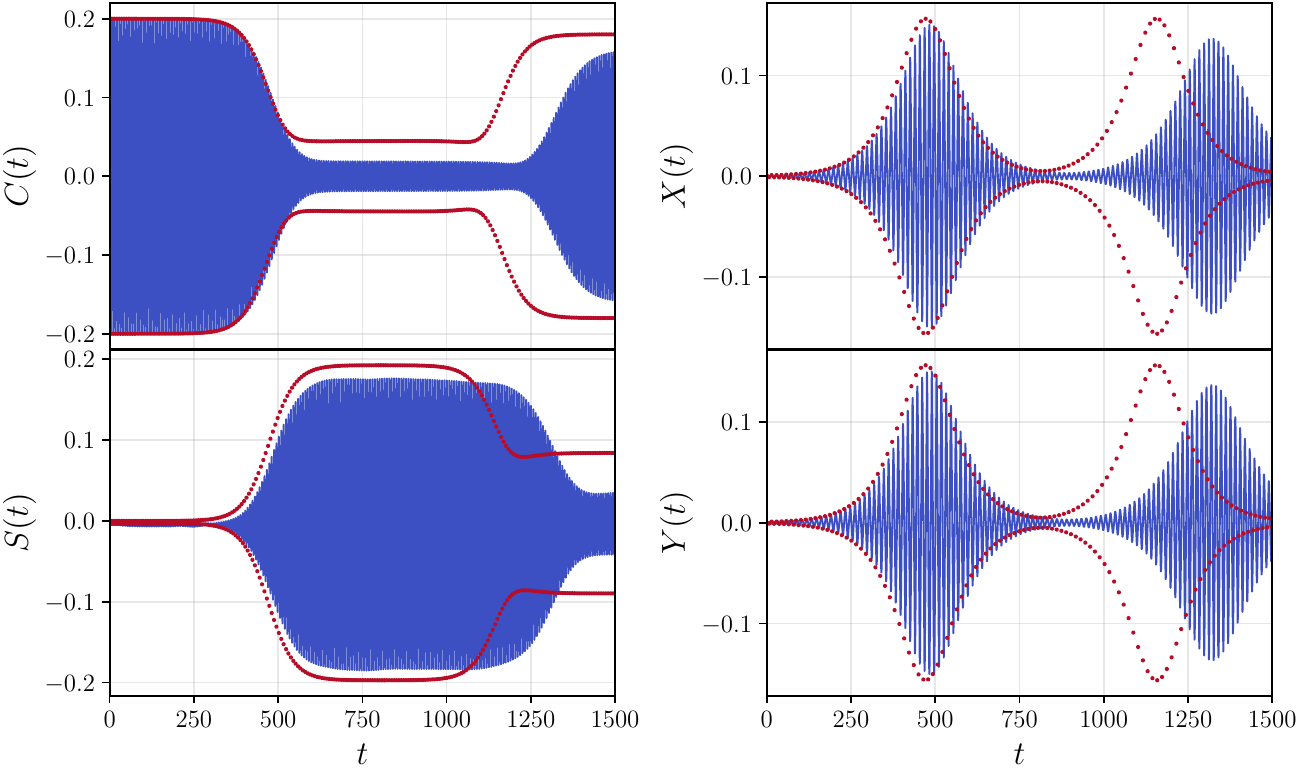}  
    \caption{\justifying
    Time evolution of the amplitudes of the vector, shape and zero modes. The blue curves show the extracted field theory amplitudes, while the red dots display the CCM predictions.}
    \label{fig: CCM vs FT}
\end{figure}

These expectations are confirmed in Fig. \ref{fig: CCM vs FT}, where the time evolution of the different amplitudes is shown. We can clearly see the consecutive resonant energy transfers in the predicted order which is followed, at later times, by the same sequence of resonances in the opposite direction. Additionally, the plotted CCM results (red dots) seem to be in good agreement with the numerical simulations up to the first oscillation of the zero mode. The deviations that emerge at later times in the zero-mode modulation can be attributed to energy loss via radiation. In particular, the second peak of the zero-mode modulation as well as the amplitude of the vector mode are clearly reduced in the field theory simulations, while in the CCM the peak amplitude remains unchanged. 

\begin{figure}[h]   
    \centering
    \includegraphics[width=\textwidth]{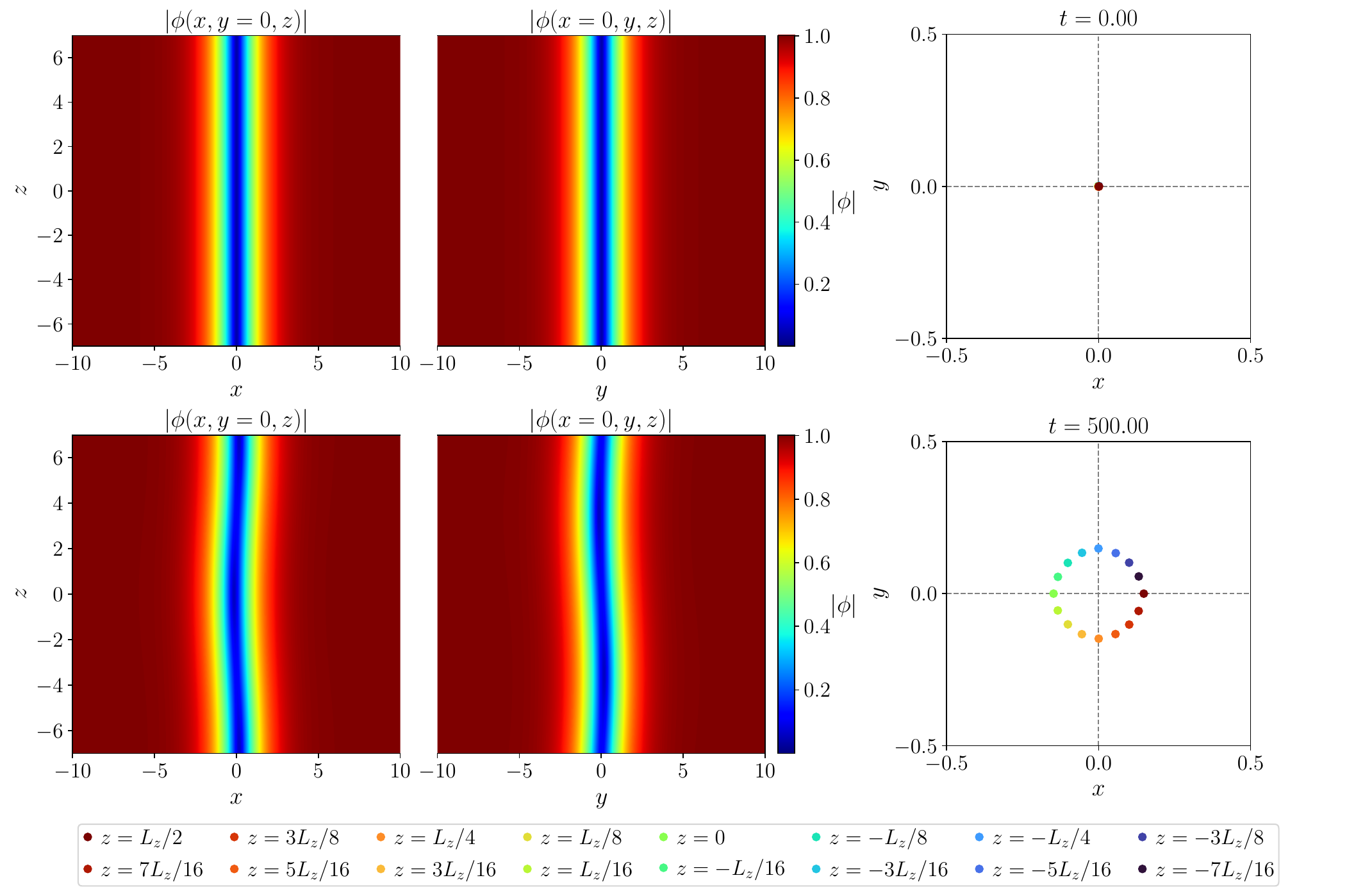}  

    \caption{\justifying
    Snapshots of the string configuration before and during the parametric resonance. The left and middle panels show the modulus of the scalar field in the $x - z$ and $y - z$ planes, respectively, while the right panels display the reconstructed string core projected onto the $x - y$ plane, with each point corresponding to a different value of $z$. The upper row shows the initially straight string at $t=0$, whereas the lower row shows the helicoidal transverse deformation developed at $t=500.0$. Additional snapshots of the evolution are provided in Appendix~\ref{App:5}, while the corresponding videos can be found in Ref.~\cite{ancillary_videos}.}
    \label{fig: snapshots_second_case}
\end{figure}

In \cref{fig: snapshots_second_case} we show the modulus of the Higgs field at $t=0$ and $t=555.80$, corresponding to the maximum of the zero-mode oscillation. Owing to the symmetric initial conditions, the two transverse amplitudes will remain equal throughout the evolution, as seen in \cref{fig: CCM vs FT}. As a result, the string will form a circular helix around its original axis, with a time-dependent radius that reaches its maximum close to the instant shown in \cref{fig: snapshots_second_case}.

Finally, we would like to emphasize some aspects of the phenomenon described above. Although we have restricted the presented numerical analysis to the BPS case, we note that the instability between the vector and transverse zero modes is a general feature of the model, independent of the chosen $\lambda$, as the principal instability condition remains the same and can be achieved for all parameter regions, $k_0=\omega_z(\lambda)/2$. Hence, it can be explored just by adjusting the longitudinal size of the simulation box so that one of the allowed wavenumbers satisfies the condition.

Second, the subsequent driven excitation of the shape mode, starting from an initially excited vector mode, is particularly easy to realize in the BPS case thanks to the degeneracy of the massive modes. Away from this limit, however, their frequencies are different, so the instability between vector and zero modes can still occur, but the subsequent growth in the shape mode's amplitude will generically be absent as a leading order resonance. We emphasize that, although the excitation of the shape mode's amplitude is an interesting addition to the findings in the section, the main result of this section is the instability of the transverse zero modes induced by the vector mode.


\section{Conclusions}
\label{conclusions}

In this work, we have investigated the excitation spectrum of local strings in the Abelian–Higgs model, with particular emphasis on vector modes. This internal degree of freedom originates from gauge field fluctuations localized within the vortex core, where the suppression of the Higgs condensate reduces the effective gauge field mass relative to its asymptotic value in the vacuum.

We have provided a detailed quantitative characterization of these modes and determined the regions of parameter space in which they arise. Physically, the vector excitations correspond to oscillatory helical magnetic field as well as longitudinal electric field configurations propagating along the string. Their structure closely resembles that of transverse magnetic modes in cylindrical electromagnetic waveguides, with the vortex core acting as a smooth waveguide for the gauge field.

As with other localized internal excitations, the vector modes decay through nonlinear couplings to radiative degrees of freedom, by mechanisms closely analogous to those governing the relaxation of the scalar shape mode.

For sufficiently small values of ($\lambda$), the dominant perturbative decay channel arises from a quadratic coupling to radiation. This interaction produces a power law decrease in both the amplitude and the energy of the localized excitation.

A central result of our analysis is that, in contrast to the shape mode, the vector mode does not merge with the continuum as ($\lambda$) is increased. Consequently, for ($\lambda>1.5$), beyond the point at which the shape mode enters the continuum, the vector mode constitutes the only discrete internal excitation supported by the string and is therefore expected to play a dominant role in its microscopic dynamics. Moreover, for ($\lambda>3.8$), the leading-order radiative channel becomes kinematically inaccessible. Considered in isolation, this result might suggest that the vector excitation becomes substantially longer lived above this threshold. The presence of quasinormal modes, however, qualitatively modifies this conclusion.

The relevant quasinormal modes contain a localized scalar component together with a nonlocalized vector component. When the vector bound state is excited with frequency ($\omega$), nonlinear interactions can source a quasinormal mode at the second harmonic, ($2\omega$). The extended vector component of this quasinormal excitation then transports energy away from the vortex and radiates it to spatial infinity. This process provides an additional decay channel for the localized vector mode, even when direct scalar radiation is forbidden at the lowest perturbative order.

A further important property of the vector mode is its coupling to the translational zero modes of the string. Using fully nonlinear field theory simulations of strings initially excited in the localized vector mode, we have demonstrated the existence of a parametric instability involving these two sectors. We have also formulated an effective collective coordinate model that reproduces the essential features of this interaction. The instability enables energy transfer from the vector excitation to the translational modes and, conversely, from the translational dynamics back into the internal degree of freedom. These results support the conjecture advanced in Ref.~\cite{Aurrekoetxea:2026xrl} that such parametric instabilities may represent a generic feature of internal excitations in extended nonlinear objects. Closely related phenomena have also been observed in the dynamics of global strings \cite{Blanco-Pillado:2022axf} as well as the shape mode in the local string~\cite{Aurrekoetxea:2026xrl}.

A natural direction for future work is the construction of a systematic effective field theory of Abelian–Higgs strings that incorporates the vector mode as a massive degree of freedom propagating on the string worldsheet, extending the approaches developed in Refs.~\cite{Aurrekoetxea:2026xrl}. For the scalar shape mode, the leading interaction with the worldsheet geometry takes the form ($\theta\mathcal{R}$), where ($\theta$) denotes the corresponding worldsheet amplitude for the internal degree of freedom and ($\mathcal{R}$) is the worldsheet Ricci scalar. The results presented here indicate that the coupling of the vector mode must have a qualitatively different tensorial and symmetry structure. Determining the symmetry allowed interactions between this excitation and the worldsheet geometry, together with its direct coupling to the scalar shape mode, will be essential for constructing a complete low energy effective description of excited Abelian–Higgs strings. These questions are currently under investigation and we hope to report on them soon.

\acknowledgments

J.J.B.P. would like to thank A. Alonso-Izquierdo, K. D. Olum and A. Vilenkin 
for a number of useful conversations on this topic over the years.
This work has been supported in part by the PID2021-123703NB-
C21 and PID2024-156016NB-I00 grants funded by
MCIN/AEI/10.13039/501100011033/and by ERDF;“ A
way of making Europe”; the Basque Government grant
(IT-1628-22) and the Basque Foundation for Science
(IKERBASQUE). JQ has been supported also in part by Spanish Ministerio de Ciencia e Innovación (MCIN) with funding from the grant PID2023-148409NB-I00 MTM.


\appendix

\section{Quantum mechanical SUSY structure in the BPS limit}
\label{App.1}
As mentioned in the main part of the text, the level crossing between the shape and vector modes is due to the hidden QM SUSY structure of the theory arising in the BPS case. In this limit, the static equations for the background vortex string profiles are given by the first order Bogomolnyi equations
\begin{equation}
    f'(r)=\frac{1-a(r)}{r}f(r)\,,\qquad \frac{a'(r)}{r}=\frac{1-f(r)^2}{2}\,.
\end{equation}

On the other hand, the second-order fluctuation operator associated with the shape mode perturbations $\left(\varphi(r)\,,\,\alpha(r)\right)^T$, the one that gives rise to the spectral problem (\ref{eq: shape_mode_spec_problem}), can be written as
\begin{equation}
    \mathcal{H}^+=
    \begin{pmatrix}
        -\partial_r^2-\frac{1}{r}\partial_r+\frac{1}{2}\left(3f(r)^2-1\right)+\frac{(1-a(r))^2}{r^2} & -\frac{2f(r)}{r}(1-a(r))\\
        -\frac{2f(r)}{r}(1-a(r)) & -\partial_r^2-\frac{1}{r}\partial_r+f(r)^2+\frac{1}{r^2}
    \end{pmatrix}\,.
\end{equation}

In the BPS limit, this operator admits a factorization in terms of first order operators, $\mathcal{H}^+=\mathcal{D}^\dagger\mathcal{D}$, with
\begin{equation}
    \mathcal{D}=
    \begin{pmatrix}
        \partial_r-\frac{1-a(r)}{r} & f(r)\\
        f(r) & \partial_r+\frac{1}{r}
    \end{pmatrix}
    \quad\rm{and}\quad
    \mathcal{D}^\dagger=
    \begin{pmatrix}
        -\partial_r-\frac{2-a(r)}{r} & f(r)\\
        f(r) & -\partial_r
    \end{pmatrix}\,.
\end{equation}

Equivalently, by reversing the order of $\mathcal{D}$ and $\mathcal{D}^\dagger$, we obtain
\begin{equation}
    \mathcal{H}^-=\mathcal{D}\mathcal{D}^\dagger=
    \begin{pmatrix}
        -\partial^2_r-\frac{1}{r}\partial_r+\frac{(2-a(r))^2}{r^2}+\frac{1+f(r)^2}{2} & 0\\
        0 & -\partial_r^2-\frac{1}{r}\partial_r+f(r)^2
    \end{pmatrix}\,,
\end{equation}
which is normally referred to as the supersymmetric quantum mechanical partner of $\mathcal{H}^+$. As we can see, the $(2,2)$ element of this diagonal partner of $\mathcal{H}^+$ is precisely the radial operator acting on the vector perturbation in equation (\ref{eq: eigenproblem vector mode}). Therefore, a configuration of the form $(0\,,\,\psi(r))^T$ will be an eigenfunction of $\mathcal{H}^-$. 

The fact that we can decompose these second-order operators into $\mathcal{D}$ and $\mathcal{D}^\dagger$ allows us to embed both of them in a supersymmetric QM system given by the ``enlarged" hamiltonian 
\begin{equation}
    \mathcal{H}=
    \begin{pmatrix}
        \mathcal{H}^+ & 0\\
        0 & \mathcal{H}^-
    \end{pmatrix}\,,\qquad \mathcal{H}\xi=\Omega^2\xi\,,
\end{equation}
where $\xi$ is the state containing the eigenfunctions of both $\mathcal{H}^+$ and $\mathcal{H}^-$. The operators that map states between these two sectors are the supercharges
\begin{equation}
    \mathcal{Q}=
    \begin{pmatrix}
        0&0\\
        \mathcal{D}&0
    \end{pmatrix}
    \quad\rm{and}
    \quad\mathcal{Q}^\dagger=
    \begin{pmatrix}
        0&\mathcal{D}^\dagger\\
        0&0
    \end{pmatrix}\,.
\end{equation}

In conjunction with $\mathcal{H}$, the supercharges satisfy the superalgebra
\begin{align}
    [\mathcal{H},\mathcal{Q}]&=[\mathcal{H},\mathcal{Q}^\dagger]=0\,,\\
    \{\mathcal{Q},\mathcal{Q}^\dagger\}&=\mathcal{H}\,,\\
    \{\mathcal{Q},\mathcal{Q}\}&=\{\mathcal{Q}^\dagger,\mathcal{Q}^\dagger\}=0\,.
\end{align}

Let us consider the spectral problem for the vector mode
\begin{equation}
    \mathcal{H}
    \begin{pmatrix}
        0\\
        \xi^-
    \end{pmatrix}
    =\Omega^2
    \begin{pmatrix}
        0\\
        \xi^-
    \end{pmatrix}\,,
    \qquad\xi^-(r)=
    \begin{pmatrix}
        0\\
        \psi(r)
    \end{pmatrix}\,.
\end{equation}

 We can now apply the adjoint supercharge operator in both sides
\begin{equation}
    \mathcal{Q}^\dagger\mathcal{H}
    \begin{pmatrix}
        0\\
        \xi^-
    \end{pmatrix}
    =\Omega^2\mathcal{Q}^\dagger
    \begin{pmatrix}
        0\\
        \xi^-
    \end{pmatrix}\,
    \Longleftrightarrow \,
    \mathcal{H}^+(\mathcal{D}^\dagger\xi^-)=\Omega^2(\mathcal{D}^\dagger\xi^-)
\end{equation}

This shows that $\mathcal{Q}^\dagger$ maps the vector mode embedded in $\xi^-$ into an eigenstate of the $\mathcal{H}^+$ sector, namely the corresponding shape mode state. Thus, not only vector and shape modes share the same nonzero eigenvalues in the BPS limit, but their profiles can be related as follows
\begin{equation}
    \begin{pmatrix}
        \varphi(r)\\
        \alpha(r)
    \end{pmatrix}
    =\frac{1}{\Omega}\mathcal{D}^\dagger
    \begin{pmatrix}
        0\\
        \psi(r)
    \end{pmatrix}
    =\frac{1}{\Omega}
    \begin{pmatrix}
        f(r)\psi(r)\\
        -\psi'(r)
    \end{pmatrix}\,,
\end{equation}
where $\Omega^{-1}$ is the normalization factor.

In the language of \cite{Alonso-Izquierdo:2015tta}, vector modes correspond to the so-called Class A $\mathcal{H^-}$ eigenmodes, whereas solutions to the eigenvalue problem given by the (1,1) element of $\mathcal{H}^-$ would correspond to Class B $\mathcal{H^-}$ eigenmodes. As shown there, bound states of the latter eigenproblem do not exist. The identification of the Class A sector with the vector mode is particularly relevant for the $3+1$ theory. Indeed, in the $2+1$ vortex setup, this sector appears only as an auxiliary partner in the QM SUSY framework, whereas once the vortex is lifted to a string, the same operator is associated with a physical longitudinal fluctuation of the gauge field.

\section{Number of bound states as $\lambda\to$0}
\label{App.2}

We want to check how the number of bound states coming from the eigenproblem (\ref{eq: eigenproblem vector mode}) changes as $\lambda\to 0$. The width of the scalar core of the vortex is approximately $\delta_\phi\sim\frac{1}{\sqrt{\lambda}}$. At first glance, the fact that the width of this soliton tends to infinity as $\lambda\to0$ is already a good enough intuitive explanation of why we should expect an arbitrarily large number of bound states in that limit. To make the previous statement more rigorous, we can make use of the Sturm comparison theorem:

\begin{theorem}
    Let $u_1$ and $u_2$ be nontrivial solutions of 
    \begin{align}
        u_1''+q_1(x)u_1=0\,,\\
        u_2''+q_2(x)u_2=0\,,
    \end{align}
    on an interval $I$, and suppose $q_1(x)\geq q_2(x)$ for all $x\in I$. Then, $u_1(x)$ has at least one zero between every two consecutive zeros of $u_2(x)$ (unless $q_1(x)\equiv q_2(x)$ and $u_1(x)$ is a constant multiple of $u_2(x)$).
\end{theorem}

It is well-known that for a Sturm-Liouville eigenvalue problem, the eigenfunctions will be determined by their number of zeros, in the sense that the $n$-th eigenfunction has exactly $n$ nodes. Therefore, counting zeros is a direct way to infer the number of bound states the problem hosts.

We can convert our eigenvalue problem at hand to a one-dimensional Schrödinger problem with the variable change $\eta(r)=\psi(r)\sqrt{r}$. Then, the differential equation reads
\begin{equation}\label{eq: Schr_eq_vector}
\eta''(r)+U_1(r)\eta(r)=0,\quad U_1(r)=\left(\omega^2+\frac{1}{4r^2}-f(r)^2\right)\,.
\end{equation}

Hence, if we are able to find an analogous problem
\begin{equation}\label{eq: Schr_eq_comparison}
\eta''(r)+U_2(r)\eta(r)=0,\quad U_2(r)=\left(\omega^2+\frac{1}{4r^2}-V(r)\right)\,,
\end{equation}
with $U_2(r)\leq U_1(r)$, or equivalently, $V(r)\geq f(r)^2$, Sturm's comparison theorem guarantees that between two consecutive zeros of a solution to the eigenproblem (\ref{eq: Schr_eq_comparison}) there is at least one zero of the corresponding solution of (\ref{eq: Schr_eq_vector}). Consequently, if the eigenvalue problem (\ref{eq: Schr_eq_comparison}) supports $N$ bound states, our original problem will support at least $N-1$ bound states.
Let that potential $V(r)$ be
\begin{equation}
    V(r)=\begin{cases}
        c\geq f(r)^2\,,&\quad r<\delta_\phi\,,\\
        1>c\,,&\quad r>\delta_\phi\,.
    \end{cases}
\end{equation}

We can now solve our Sturm-Liouville equation for $V(r)$ in both regions and apply matching conditions to get a transcendental equation at $r=\delta_\phi$, whose zeros in the $\omega\in(\sqrt{c}, 1)$ region should tell us the amount of bound states that this new potential hosts.

In the inner region, $r<\delta_\phi$, the equation becomes
\begin{equation}
    -\psi''(r)-\frac{\psi'(r)}{r}+c\psi(r)=\omega^2\psi(r)\,,
\end{equation}
whose solution regular at the origin is
\begin{equation}
    \psi_{\rm in}(r)=J_0\left(r\sqrt{\omega^2-c}\right)\,.
\end{equation}

In the outside region $r>\delta_\phi$, we have
\begin{equation}
    -\psi''(r)-\frac{\psi'(r)}{r}+\psi(r)=\omega^2\psi(r)\,,
\end{equation}
and the solution that decays at infinity is
\begin{equation}
    \psi_{\rm out}(r)=K_0\left(r\sqrt{1-\omega^2}\right)\,.
\end{equation}

We can now apply matching conditions at $r=\delta_\phi$, that is, 
\begin{equation}
    \psi_{\rm in}(r)\big|_{r=\delta_\phi}=\psi_{\rm out}(r)\big|_{r=\delta_\phi}\,,\qquad\psi'_{\rm in}(r)\big|_{r=\delta_\phi}=\psi'_{\rm out}(r)\big|_{r=\delta_\phi}\,.
\end{equation}

They can be combined into
\begin{equation}
    \frac{\psi'_{\rm in}(r)\big|_{r=\delta_\phi}}{\psi_{\rm in}(r)\big|_{r=\delta_\phi}}=\frac{\psi'_{\rm out}(r)\big|_{r=\delta_\phi}}{\psi_{\rm out}(r)\big|_{r=\delta_\phi}}\,,
\end{equation}
which leads to the transcendental equation
\begin{equation}
    \sqrt{\omega^2-c} \frac{J_1\left(\frac{\sqrt{\omega^2-c}}{\sqrt{\lambda }}\right)}{J_0\left(\frac{\sqrt{\omega^2-c}}{\sqrt{\lambda }}\right)}=\sqrt{1-\omega^2}\frac{ K_1\left(\frac{\sqrt{1-\omega^2}}{\sqrt{\lambda }}\right)}{K_0\left(\frac{\sqrt{1-\omega^2}}{\sqrt{\lambda }}\right)}\,.
\end{equation}

This equation can be further simplified by using the limiting expressions for the Bessel functions in the desired limit, leading to
\begin{equation}
    \cot \left(\frac{\sqrt{\omega^2-c}}{\sqrt{\lambda }}+\frac{\pi }{4}\right)=-\frac{\sqrt{1-\omega^2}}{\sqrt{\omega^2-c}}\,.
\end{equation}

Each time the cotangent crosses the r.h.s. of the previous equation, we get a bound state (in the region $\omega\in(\sqrt{c},1)$), and this will occur once in every oscillation of the cotangent function. Since the period of the cotangent is $\pi$ and the phase difference within the allowed range of $\omega$ is $\Delta=\sqrt{\frac{1-c}{\lambda}}$, the number of oscillations is
\begin{equation}
    N\sim\frac{\Delta}{\pi}=\frac{1}{\pi}\sqrt{\frac{1-c}{\lambda}}\xrightarrow{\lambda\to0}\infty\,.
\end{equation}

Thus, as $\lambda\to0$, the number of bound states of the comparison problem becomes arbitrarily large, and the same is therefore true for our original problem.

\section{Numerical details}
\label{App.3}

\subsection{1+1 dimensions}

In section 3, the cylindrical symmetry of the problem at hand has allowed us to reduce system's dynamics to the following set of radial equations:
\begin{align}\label{eq:red_1}
\frac{\partial^2F}{\partial t^2}-\frac{\partial^2F}{\partial r^2}-\frac{1}{r}\frac{\partial F}{\partial r}+\frac{1}{r^2}(1-B)^2F+FA_z^2-\frac{\lambda}{2}(1-F^2)F&=0\,,\\
\label{eq:red_2}
\frac{\partial^2B}{\partial t^2}-\frac{\partial^2B}{\partial r^2}+\frac{
1}{r}\frac{\partial B}{\partial r}-(1-B)F^2&=0\,,\\
\label{eq:red_3}
\frac{\partial^2A_z}{\partial t^2}-\frac{\partial^2A_z}{\partial r^2}-\frac{1}{r}\frac{\partial A_z}{\partial r}+A_zF^2&=0\,,
\end{align} 

where $F(r,t)$ is the radial profile of the scalar field, $B(r,t)=rA_\theta(r,t)$ is the (rescaled) angular component of the gauge field, and $A_z(r,t)$ its axial component. Notice that in this setting, the Gauss' law is automatically satisfied, ensuring that the system evolves perpendicularly to the gauge orbit, and hence there is no need to monitor it.

These equations have been solved employing the staggered leapfrog method, second order accurate in time, combined with a second order finite-difference discretization for the spatial derivatives, in a simulation box of size $[0,L]$, where $L=70.0$ with a uniform spacing $\Delta r=0.01$, and $\Delta t=0.001$.

In order to avoid radiation being reflected back at the edge of the box, absorbing boundary conditions have been employed at $r=L$ \cite{Mur4091495}. Absorbing boundary conditions come in handy as they allow us to annihilate outgoing radial radiation of the form

\begin{equation}
    u(r,t)=\frac{\cos(\omega t -kr +\delta)}{\sqrt{r}}\,,
\end{equation}
by imposing the following condition at the boundary,
\begin{equation}
    \partial_t u(r,t)+\partial_r u(r,t)+\frac{1}{2r}u(r,t)=0\,.
\end{equation}

The previous condition can be further refined if we know the nature of the outgoing radiation. Indeed, in section 3, where scalar and gauge (angular component) radiation is dominated at leading order by an angular frequency of $\omega=2\omega_z$ and their wavenumbers $k_{\phi,\theta}$ are known, the condition can be tuned as
\begin{equation}
    \partial_t R_{\phi,\theta}(r,t)+\frac{2\omega_z}{k_{\phi,\theta}}\partial_r R_{\phi,\theta}+\frac{2\omega_z}{2k_{\phi,\theta}r}R_{\phi,\theta}(r,t)=0\,.
\end{equation}

For further safety, the absorbing boundary conditions have been complemented with a damping layer close to the boundary. This amounts to adding first order time derivatives of the form $-\epsilon(r)\partial_t$ at the r.h.s. of equations (\ref{eq:red_1}) - (\ref{eq:red_3}), with
\begin{equation}
\epsilon(r)=
\begin{cases}
\displaystyle 0, 
& \displaystyle r<r_*=\frac{5L}{6}\,,\\[8pt]
\displaystyle \left(\frac{r-r_*}{10}\right)^4, 
& \displaystyle r_*<r<L\,.
\end{cases}
\end{equation}

The agreement between the results from this numerical scheme and the semianalytic procedure, as seen in the main part of the text, has been remarkable. However, we must note that for values of $\lambda$ in the range $3.5\lesssim\lambda\lesssim3.8$, i.e., close to the threshold where the scalar radiation channel closes (at leading, nonlinear order in $C(t)$, that is), there has been a slight but noticeable departure from the expected power law formula. We attribute this difference to the fact that in this interval $k_\phi$ reaches its smallest values and, thus, the wavelength of the scalar radiation reaches scales comparable to the size of the box, potentially leading to some undesired, pathological interactions. Nevertheless, by increasing the size of the box as well as widening the damping layer, the agreement with the expected decay is restored, as shown in the following Figure.

\begin{figure}[h]   
    \centering
    \includegraphics[width=1.0\textwidth]{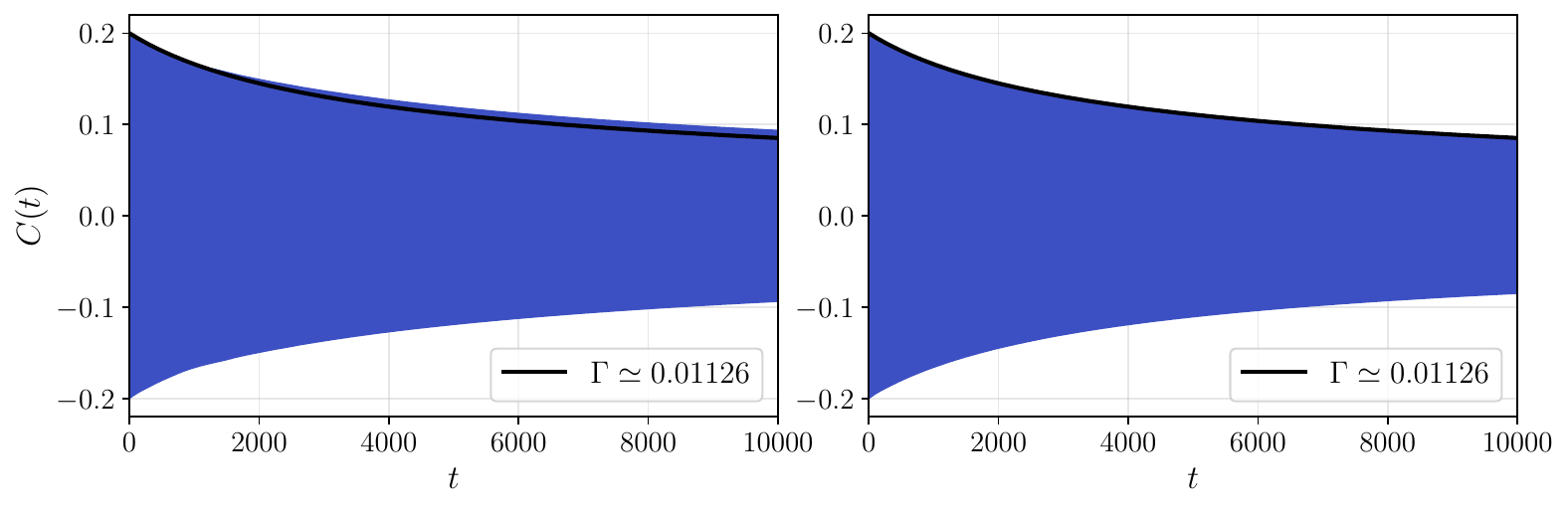}  
    \caption{\justifying
    Amplitude of the vector mode, $C(t)$, as a function of time extracted from the field theory simulations for $\lambda=3.74$. Left: length of the simulation box $L=70.0$ and damping from $r=5L/6$ onward. Right: length of the simulation box $L=150.0$ and damping from $r=2L/3$ onward. The black curves correspond to the analytical decay, with $\Gamma$ being the decay constant of the power law decay formula (\ref{eq: vector_decay})}
    \label{fig: L_comparison}
\end{figure}

\subsection{3+1 dimensions}

In this case, the setup is no longer reduced by exploiting any symmetry of the system, and the vortex string is no longer pinned down to the origin, allowing for a wider range of dynamics.

The set of equations that needs to be solved in the 3D cartesian grid is
\begin{align}
\partial_t^2\phi_1&=\partial_i\partial_i\phi_1+\phi_2\partial_iA_i+2A_i\partial_i\phi_2-\phi_1A_iA_i+\frac{\lambda}{2}\left[1-(\phi_1^2+\phi_2^2)\right]\phi_1\,,\\
\partial_t^2\phi_2&=\partial_i\partial_i\phi_2-\phi_1\partial_iA_i-2A_i\partial_i\phi_1-\phi_2A_iA_i+\frac{\lambda}{2}\left[1-(\phi_1^2+\phi_2^2)\right]\phi_2\,,\\
\partial_t^2A_j&=\partial_i\partial_iA_j-\partial_j(\partial_iA_i)+(\phi_1\partial_j\phi_2-\phi_2\partial_j\phi_1)-A_j(\phi_1^2+\phi_2^2)\,,
\end{align}
where $i,j=x,y,z$ and the scalar field has been decomposed into its real and imaginary parts, $\phi=\phi_1+i\phi_2$. On the other hand, the Gauss' law reads
\begin{equation}
    \partial_t(\partial_iA_i)=\phi_1\partial_t\phi_2-\phi_2\partial_t\phi_1\,.
\end{equation}

Once again, these equations have been solved using the staggered leapfrog method, second-order accurate in time, together with second-order finite differences for spatial derivatives. The transverse size of the 3D box was chosen to be $L_x=L_y=60$, with lattice spacings $dx=dy=0.1$ and time step $dt=0.05$. We note that simulations performed with larger transverse box sizes have not yielded appreciable differences in the results. The values of $L_z$, $dz$, and therefore $N_z=L_z/dz$ have been chosen to realize the desired longitudinal wavenumber while maintaining adequate resolution along the string. For all the simulations reported in \cref{Sec: interac}, $L_z=14.25$ and $dz=0.125$.

Regarding boundary conditions, periodic BCs have been imposed along the longitudinal $z$ direction, while in the transverse directions absorbing boundary conditions have been used.

The time-dependent amplitudes of the different modes have been extracted by first subtracting the background field profiles from the evolved fields and then projecting the resulting perturbations onto the corresponding mode profiles. The background and mode profiles, originally obtained on a one-dimensional radial grid, have been interpolated onto the 3D cartesian lattice and expressed in terms of their (Cartesian) field components before carrying out these projections and, in general, before starting the simulation.

At each time step, the (normalized) amplitudes have been obtained from the following projections. Let us first define
\begin{equation}
    \Delta\phi_a\equiv \phi_a^{\rm sim}(\mathbf{x},t)-\phi_a^{\rm bg}(\mathbf{x})\,,\qquad\Delta A_i\equiv A_i^{\rm sim}(\mathbf{x},t)-A_i^{\rm bg}(\mathbf{x})\,.
\end{equation}

Then the vector mode amplitude is extracted as
\begin{equation}
    C(t)=\frac{\int_Vd^3x\,\Delta A_z\,\psi(\mathbf{x})}{\int_Vd^3x\,\psi(\mathbf{x})^2}\,,\qquad A_z^{\rm bg}(\mathbf{x})=0\,.
\end{equation}

For the homogeneous shape mode
\begin{equation}
    S(t)=\frac{\int_Vd^3x\,[\Delta\phi_1\,\varphi_1(\mathbf{x})+\Delta\phi_2\,\varphi_2(\mathbf{x})+\Delta A_x\,\alpha_x(\mathbf{x})+\Delta A_y\,\alpha_y(\mathbf{x})]}{\int_Vd^3x\,[\varphi_1(\mathbf{x})^2+\varphi_2(\mathbf{x})^2+\alpha_x(\mathbf{x})^2+\alpha_y(\mathbf{x})^2]}\,,
\end{equation}
where, $\varphi_1(\mathbf{x})$ and $\varphi_2(\mathbf{x})$  denote the two Cartesian components of the scalar shape mode profile, while $\alpha_x(\mathbf{x})$ and $\alpha_y(\mathbf{x})$ correspond to the Cartesian components of its gauge field profile.

Finally, for the zero mode in $x$,
\begin{equation}
    X(t)=\frac{\int_Vd^3x\, \cos(k_0z)\,\left[\Delta\phi_1\,\delta_x\phi_1(\mathbf{x})+\Delta\phi_2\,\delta_x\phi_2(\mathbf{x})+\Delta A_i\,\delta_xA_i(\mathbf{x})\right]}{\int_Vd^3x\, \cos^2(k_0z)\,\left[\delta_x\phi_1(\mathbf{x})^2+\delta_x\phi_2(\mathbf{x})^2+\delta_xA_i(\mathbf{x})\,\delta_xA_i(\mathbf{x})\right]}\,,
\end{equation}
and similarly for $Y(t)$.

Let us finally mention that the Gauss' law has also been monitored during the simulations every few time steps. In all the simulations performed, the initial string configuration satisfied this constraint. Two main quantities have been checked: the maximum deviation of the constraint and its location within the box, together with the root mean square of the deviation over the entire lattice, computed as
\begin{equation}
    \mathcal{G}_{\rm rms}=\sqrt{\frac{1}{V}\int_Vd^3x\,[\partial_t(\partial_iA_i)-\phi_1\partial_t\phi_2+\phi_2\partial_t\phi_1]^2}\,.
\end{equation}

No significant growth of $\mathcal{G}_{\rm rms}$ has been observed during the simulations. Moreover, most of the times, the maximum deviations have been found to be located at the boundaries of the box, providing further confidence that they did not notably affect the dynamics of the string core and its vicinity.

\section{Integrals for the effective models}
\label{App.4}

This appendix serves as a collection of the coefficients entering the effective Lagrangians (\ref{eq: EffLag2DoF}) and (\ref{eq: EffLag4DoF}) introduced in \cref{Sec: interac}.
\subsection{Effective model with 2 DoF}

The integrals appearing in the effective Lagrangian (\ref{eq: EffLag2DoF}) are the following:
\begin{equation}
    I_{dX2}=\frac{1}{2}\int_{-L/2}^{L/2} \cos^2(k_0z) dz\int_0^{2 \pi}\int_0^\infty \left(|\delta_x\phi(r,\theta)|^2+|\delta_xA_r(r,\theta)|^2+|\delta_xA_\theta(r,\theta)|^2\right)rdrd\theta\,,
\end{equation}

\begin{equation}
    I_{dC2}=\frac{1}{2}\int_{-L/2}^{L/2} dz\int_0^{2\pi}d\theta\int_0^\infty\psi^2r dr\,.
\end{equation}

For the integrals that are purely quadratic in their amplitudes, it is useful to separate them into their longitudinal contribution, $I^\parallel$, which originates from derivatives with respect to $z$, and their transverse contributions, $I^\perp$, which is directly related to the second order transverse fluctuation operator. For the zero mode
\begin{equation}
    I_{X2}=I_{X2}^\parallel+I_{X2}^\perp\,,\qquad I_{X2}^\perp=0\,,
\end{equation}
since it is an eigenfunction of zero eigenvalue of the transverse fluctuation operator. The parallel contribution reads
\begin{equation}
    I_{X2}^{\parallel}=-\frac{k_0^2}{2}\int_{-L/2}^{L/2} \sin^2(k_0z)dz\int_0^{2\pi}\int_0^\infty \left(|\delta_x\phi|^2+|\delta_xA_r|^2+|\delta_xA_\theta|^2\right)rdrd\theta\,.
\end{equation}

For the (homogeneous) vector mode, in contrast,
\begin{equation}
    I_{C2}=I_{C2}^\parallel+I_{C2}^\perp\,,\qquad I_{C2}^\parallel=0\,,
\end{equation}
with
\begin{equation}
    I_{C2}^\perp=-\frac{1}{2}\int_{-L/2}^{L/2}dz \int_0^{2\pi}d\theta\int_0^\infty \left(f^2\psi^2+\psi'^2\right)r dr=-\frac{\omega_z^2}{2}\int dz\int_0^{2\pi}d\theta\int_0^\infty\psi^2r dr\,,
\end{equation}
where the eigenproblem of the vector mode has been used to reach the final expression. It therefore follows that
\begin{equation}
    \frac{I_{X2}}{I_{dX2}}=-k_0^2\,,\quad \frac{I_{C2}}{I_{dC2}}=-\omega_z^2\,,
\end{equation}
the former relation holds when $k_0=\frac{2\pi n}{L}$ is met.

The remaining coefficients describe the leading nonlinear self-interaction of the zero mode and its quadratic coupling to the vector mode, namely,
\begin{equation}
    I_{X4}=-\frac{1}{2}\int_{-L/2}^{L/2} \cos^4(k_0 z)dz\int_0^{2 \pi}\int_0^\infty \left[\left(|\delta_xA_r|^2+|\delta_xA_\theta|^2\right)|\delta_x\phi|^2+\frac{\lambda}{4}|\delta_x\phi|^4\right]rdrd\theta\,,
\end{equation}
and
\begin{equation}
    I_{X2C2}=-\frac{1}{2}\int_{-L/2}^{L/2} \cos^2(k_0 z)dz\int_0^{2\pi}\int_0^\infty\psi^2 |\delta_x\phi|^2r drd\theta\,.
\end{equation}

\subsection{Effective model with 4 DoF}
We now include the second zero mode, $Y(t)$, and the shape mode $S(t)$. The coefficients already defined in the 2 DoF model remain unchanged, so below we will just list the additional terms:
\begin{equation}
    I_{dY2}=\frac{1}{2}\int_{-L/2}^{L/2} \sin^2(k_0z) dz\int_0^{2 \pi}\int_0^\infty \left(|\delta_y\phi|^2+|\delta_yA_r|^2+|\delta_yA_\theta|^2\right)rdrd\theta\,,
\end{equation}

\begin{equation}
    I_{dS2}=\frac{1}{2}\int_{-L/2}^{L/2} dz \cos^2(k_s z)\int_0^{2\pi}d\theta\int_0^\infty rdr(\varphi^2+\alpha^2)\,.
\end{equation}

For the purely quadratic terms, we again separate the longitudinal and transverse contributions. For the $y$ zero mode, just as before, $I_{Y2}^\perp=0$ and
\begin{equation}
    I_{Y2}^{\parallel}=-\frac{k_0^2}{2}\int_{-L/2}^{L/2} \cos^2(k_0z)dz\int_0^{2\pi}\int_0^\infty \left(|\delta_y\phi|^2+|\delta_yA_r|^2+|\delta_yA_\theta|^2\right)rdrd\theta\,.
\end{equation}

For the shape mode,
\begin{equation}
    I_{S2}^{\parallel}=-\frac{1}{2}k_s^2\int_{-L/2}^{L/2} dz \sin^2(k_s z)\int_0^{2\pi}d\theta\int_0^\infty rdr(\varphi^2+\alpha^2)\,,
\end{equation}
and
\begin{equation}
    I_{S2}^{\perp}=-\frac{1}{2}\Omega_s^2\int_{-L/2}^{L/2} dz \cos^2(k_s z)\int_0^{2\pi}d\theta\int_0^\infty rdr(\varphi^2+\alpha^2)\,.
\end{equation}

Hence,
\begin{equation}
    \frac{I_{Y2}}{I_{dY2}}=-k_0^2\,,\quad \frac{I_{S2}}{I_{dS2}}=\frac{I_{S2}^\perp+I_{S2}^\parallel}{I_{dS2}}=-(\Omega_s^2+k_s^2)=-\omega_s^2\,.
\end{equation}

On the other hand, the nonlinear self-interaction coefficients are
\begin{align}
    I_{Y4}&=-\frac{1}{2}\int_{-L/2}^{L/2} \sin^4(k_0 z)dz\int_0^{2 \pi}\int_0^\infty \left[\left(|\delta_yA_r|^2+|\delta_yA_\theta|^2\right)|\delta_y\phi|^2+\frac{\lambda}{4}|\delta_y\phi|^4\right]rdrd\theta\,,\\
    I_{S3}&=-\frac{1}{2}\int_{-L/2}^{L/2} dz \cos^3(k_sz)\int_0^{2\pi}d\theta\int_0^\infty rdr\left(\frac{2(a-1)\varphi^2\alpha}{r}+2f\varphi\alpha^2+\lambda f\varphi^3\right)\,,\\
    I_{S4}&=-\frac{1}{2}\int_{-L/2}^{L/2} dz \cos^4(k_s z)\int_0^{2\pi}d\theta\int_0^\infty rdr\varphi^2\left(\alpha^2+\frac{\lambda}{4}\varphi^2\right)\,.
\end{align}
The cubic terms coupling different mode amplitudes are
\begin{flalign}
   \hspace{-1cm} I_{X2S}&=-\int_{-L/2}^{L/2} dz \cos^2(k_0z)\cos(k_sz)\int_0^{2\pi}d\theta\int_0^\infty rdr \Bigg[\alpha\left(\frac{a-1}{r}|\delta_x\phi|^2-\frac{(a-1)ff'}{r^2}+2ff'\cos\theta(\delta_xA_\theta)\right)\notag\\
   \hspace{-1cm} &+\varphi\left(f(\delta_xA_\theta)^2+\frac{(a-1)a'f'}{2r^2}(1+3\cos(2\theta))-\frac{f(a-1)a'}{r^3}\cos(2\theta)+\frac{\lambda}{2}f\left(|\delta_x\phi|^2+2f'^2\cos^2\theta\right)\right)\notag\\
   \hspace{-1cm} &+\varphi'\frac{f(a-1)a'\sin^2\theta}{r^2}\Bigg]\,,\\
   \hspace{-1cm} I_{Y2S}&=-\int_{-L/2}^{L/2} dz \sin^2(k_0z)\cos(k_sz)\int_0^{2\pi}d\theta\int_0^\infty rdr \Bigg[\alpha\left(\frac{a-1}{r}|\delta_y\phi|^2-\frac{(a-1)ff'}{r^2}+2ff'\sin\theta(\delta_yA_\theta)\right)\notag\\
   \hspace{-1cm} &+\varphi\left(f(\delta_yA_\theta)^2+\frac{(a-1)a'f'}{2r^2}(1-3\cos(2\theta))+\frac{f(a-1)a'}{r^3}\cos(2\theta)+\frac{\lambda}{2}f\left(|\delta_y\phi|^2+2f'^2\sin^2\theta\right)\right)\notag\\
   \hspace{-1cm} &+\varphi'\frac{f(a-1)a'\cos^2\theta}{r^2}\Bigg]\,,
\end{flalign}
\begin{flalign}
   \hspace{-1cm} I_{XYC}&=-k_0\int_{-L/2}^{L/2} dz\int_0^{2\pi}d\theta\int_0^\infty r dr\left(\frac{(a-1)ff'\psi}{r}\right)\,,\\
   \hspace{-1cm} I_{C2S}&=-\int_{-L/2}^{L/2} dz \cos(k_sz)\int_0^{2\pi}d\theta\int_0^\infty rdr f\varphi\psi^2\,.
\end{flalign}
Finally, the remaining coefficients describing quartic interactions are
\begin{align}
    I_{X2Y2}&=-\frac{1}{2}\int dz\,\cos^2(k_0z)\sin^2(k_0z)\int_0^{2\pi}d\theta\int_0^\infty rdr\Bigg[\left((\delta_xA_r)^2+(\delta_xA_\theta)^2\right)|\delta_y\phi|^2\notag\\
    &+\left((\delta_yA_r)^2+(\delta_yA_\theta)^2\right)|\delta_x\phi|^2+\frac{\lambda}{2}|\delta_x\phi|^2|\delta_y\phi|^2 +\lambda\left(\operatorname{Re}\left[(\delta_x\phi)^*\delta_y\phi\right]\right)^2\Bigg]\,,\\
    I_{Y2C2}&=-\frac{1}{2}\int_{-L/2}^{L/2} \sin^2(k_0 z)dz\int_0^{2\pi}\int_0^\infty\psi^2 |\delta_y\phi|^2r drd\theta\,,\\
    I_{X2S2}&=-\int_{-L/2}^{L/2} dz \cos^2(k_0z)\cos^2(k_sz)\int_0^{2\pi}d\theta\int_0^\infty rdr\Bigg[|\delta_x\phi|^2\left(\frac{\alpha^2}{2}+\frac{\lambda}{4}\varphi^2\right)\notag\\
    &+\frac{\varphi^2}{2}\left((\delta_xA_{r})^2+(\delta_xA_{\theta})^2\right)+\frac{2\varphi\alpha f'a'}{r}\cos^2\theta+\frac{\lambda}{2}f'^2\varphi^2\cos^2\theta\Bigg]\,,\\
    I_{Y2S2}&=-\int dz\,\sin^2(k_0z)\cos^2(k_sz)\int_0^{2\pi}d\theta\int_0^\infty rdr\Bigg[|\delta_y\phi|^2\left(\frac{\alpha^2}{2}+\frac{\lambda}{4}\varphi^2\right)\notag\\
    &+\frac{\varphi^2}{2}\left((\delta_yA_r)^2+(\delta_yA_\theta)^2\right)+\frac{2\alpha\varphi f'a'}{r}\sin^2\theta+\frac{\lambda}{2}f'^2\varphi^2\sin^2\theta\Bigg]\,,\\
    I_{C2S2}&=-\frac{1}{2}\int_{-L/2}^{L/2} dz \cos^2(k_sz)\int_0^{2\pi}d\theta\int_0^\infty rdr \varphi^2\psi^2\,.
\end{align}

\section{Additional snapshots of the parametric instability}
\label{App:5}
In this Appendix, we provide additional snapshots of the two field theory simulations discussed in Sec.~\ref{Sec: interac}. The sequences complement the representative configurations shown in Figs.~\ref{fig: snapshots_first_case} and~\ref{fig: snapshots_second_case}. 

\begin{figure}[t]   
    \centering
    \includegraphics[width=0.9\textwidth]{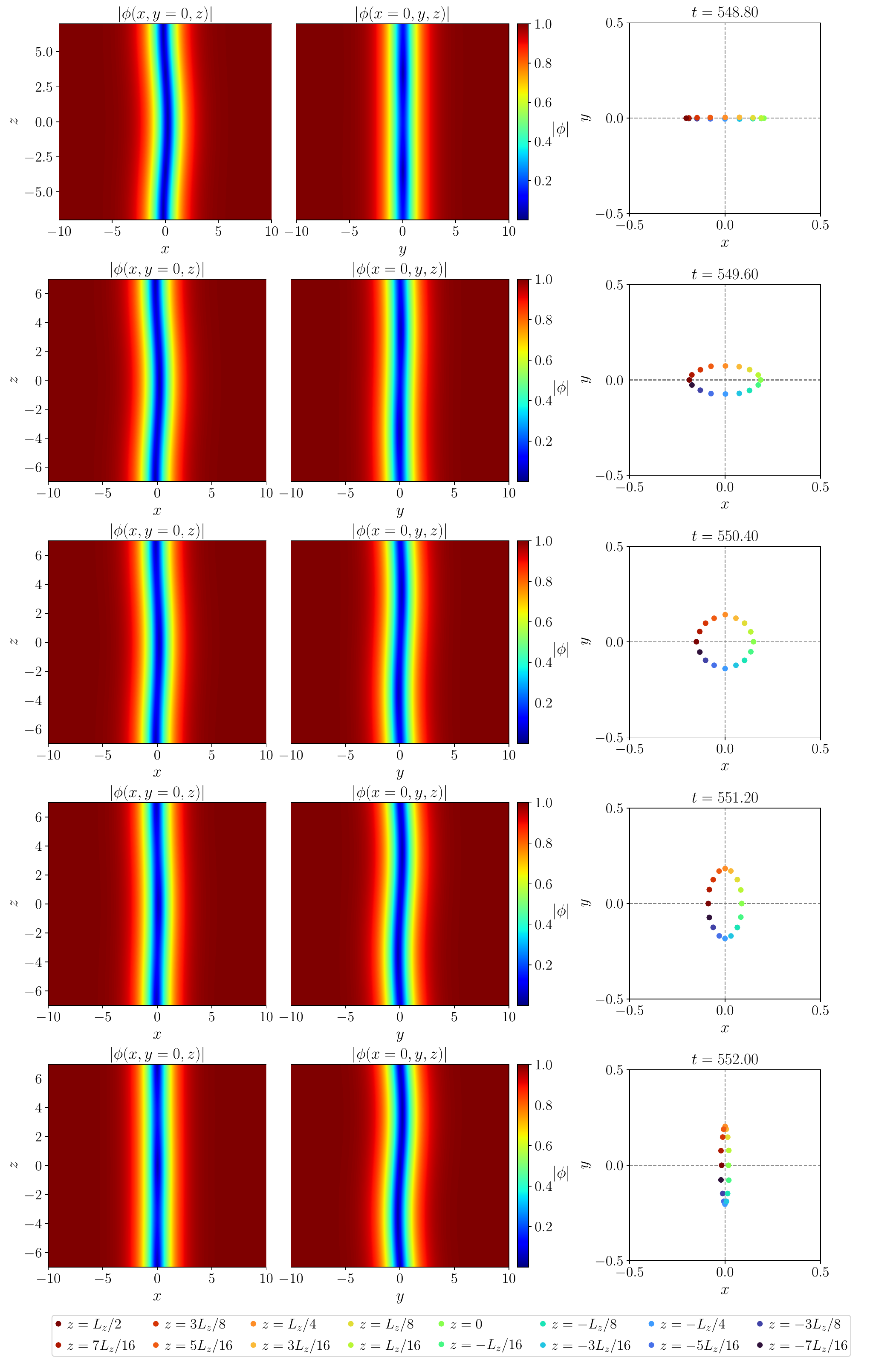}  
    \caption{\justifying 
    Complementary snapshots to Fig.~\ref{fig: snapshots_first_case}
    .}
    \label{fig:appendix_first_case}
\end{figure}

\begin{figure}[t]   
    \centering
    \includegraphics[width=0.9\textwidth]{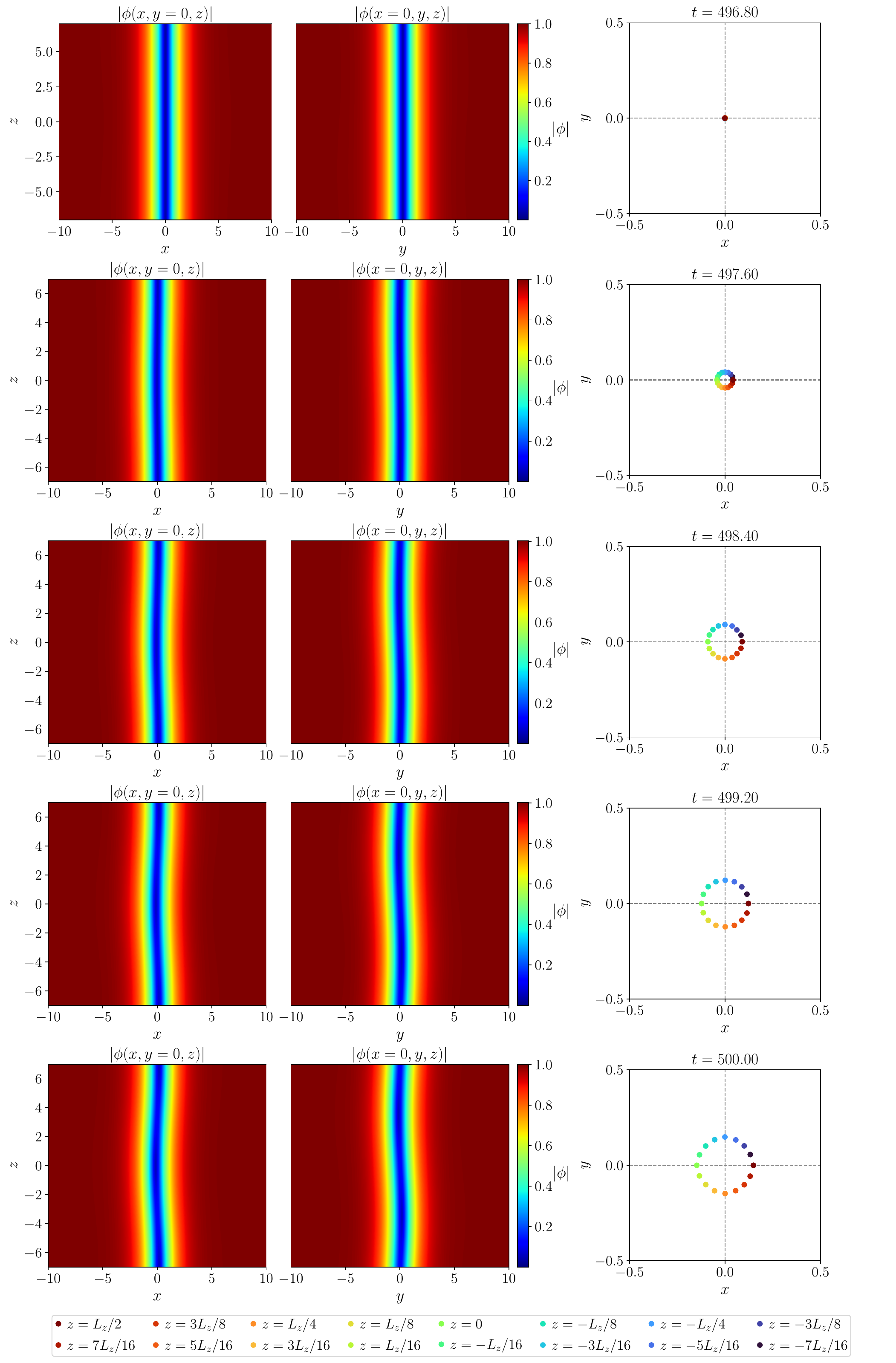}  
    \caption{\justifying
    Complementary snapshots to Fig.~\ref{fig: snapshots_second_case}.}
    \label{fig:appendix_second_case}
\end{figure}

Fig.~\ref{fig:appendix_first_case} shows the evolution for the simulation initialized with $Y(0)=0$, while Fig.~\ref{fig:appendix_second_case} corresponds to the symmetric initial conditions $X(0)=Y(0)$.  The complete time evolution of each
simulation is provided in the corresponding ancillary video.

\clearpage
\bibliography{bibliography}

\end{document}